# EEG Right & Left Voluntary Hand Movement-based Virtual Brain-Computer Interfacing Keyboard Using Hybrid Deep Learning Approach

Biplov Paneru[1], Bipul Thapa[2], Bishwash Paneru[3], Sanjog Chhetri Sapkota[4]

[1]Department of Electronics and Communication Engineering, Nepal Engineering College, Affiliated to Pokhara University, Bhaktapur, Nepal

[2]Department of Computer and Information Sciences, University of Delaware, Newark, USA

[3]Department of Applied Sciences and Chemical Engineering, Institute of Engineering Pulchowk Campus, Affiliated to Tribhuvan University, Lalitpur, Nepal

[4]Nepal Research and Collaboration Center, Nepal

***Corresponding author email***: *biplovp019402@nec.edu.np*

Abstract

Brain-machine interfaces (BMIs), particularly those based on electroencephalography (EEG), offer promising solutions for assisting individuals with motor disabilities. However, challenges in reliably interpreting EEG signals for specific tasks, such as simulating keystrokes, persist due to the complexity and variability of brain activity. Current EEG-based BMIs face limitations in adaptability, usability, and robustness, especially in applications like virtual keyboards, as traditional machine-learning models struggle to handle high-dimensional EEG data effectively. To address these gaps, we developed an EEG-based BMI system capable of accurately identifying voluntary keystrokes, specifically leveraging right and left voluntary hand movements. Using a publicly available EEG dataset, the signals were pre-processed with band-pass filtering, segmented into 22-electrode arrays, and refined into event-related potential (ERP) windows, resulting in a 19x200 feature array categorized into three classes: resting state (0), 'd' key press (1), and 'l' key press (2). Our approach employs a hybrid neural network architecture with BiGRU-Attention as the proposed model for interpreting EEG signals, achieving superior test accuracy of 90% and a mean accuracy of 91% in 10-fold stratified cross-validation. This performance outperforms traditional ML methods like Support Vector Machines (SVMs) and Naive Bayes, as well as advanced architectures such as Transformers, CNN-Transformer hybrids, and EEGNet. Finally, the BiGRU-Attention model is integrated into a real-time graphical user interface (GUI) to simulate and predict keystrokes from brain activity. Our work demonstrates how deep learning can advance EEG-based BMI systems by addressing the challenges of signal interpretation and classification. By providing a comparative analysis of multiple models and implementing a real-time application, this research highlights the feasibility and reliability of BMIs in improving accessibility for individuals with motor impairments.

1. Introduction

Electroencephalography (EEG) is a neuroimaging technique used to record electrical activity in the brain through electrodes placed on the scalp [9, 10]. EEG provides a high temporal resolution, enabling real-time monitoring of neural dynamics associated with cognitive, emotional, and motor activities. Advancements in neuroimaging technologies have opened new prospects for research in neurodegenerative diseases by enabling non-invasive investigation of the brain's structural and functional dynamics [16]. This non-invasive approach makes EEG a widely used tool for investigating brain function in clinical and research settings

Building on EEG technology, Brain-Computer Interfaces (BCIs) have emerged as transformative systems that connect the brain's electrical signals to external devices, as shown in Figure 1. BCIs enable individuals to interact with assistive technologies, such as robotic arms, wheelchairs, and virtual elements like cursors or drones, using neural commands. These interfaces are particularly crucial for individuals with motor neuron disorders and neurodegenerative diseases, as they enable the control of assistive tools such as wheelchairs, and prosthetic limbs, and virtual elements like computer cursors and drones, providing pathways for communication and independent control despite severe physical limitations [1, 25].

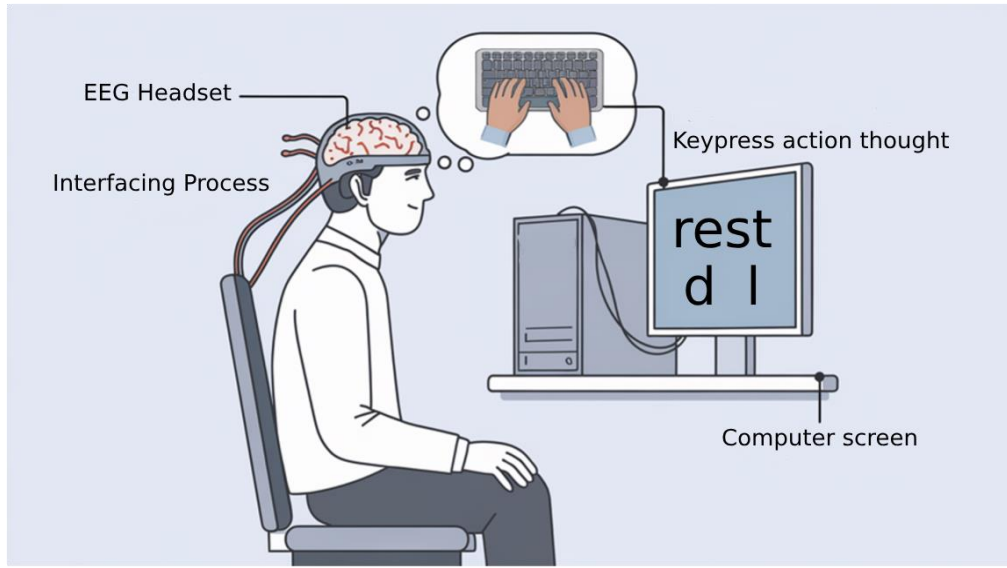

**Fig 1.** Brain-Computer Interfacing illustration

Building on these advancements, BCIs offer transformative solutions for individuals with severe motor impairments, such as those with Locked-In Syndrome (LIS). While assistive communication systems often rely on physiological signals, they frequently fall short of addressing the complex needs of these users. BCIs that use EEG-derived ERP have shown promise in real-time intent detection, enabling control of devices such as wheelchairs, prosthetics, and virtual objects like drones and cursors [2]. However, current BCI-based virtual keyboards lack adaptability and user-friendliness, which limits their accessibility for individuals with impaired motor control. These shortcomings highlight the need for more flexible and user-centred solutions in BCI systems.

The limitations of existing BCI-based virtual keyboards extend beyond usability challenges to significant technological constraints. Many current systems fail to effectively process the complex and high-dimensional nature of EEG signals, primarily due to their reliance on traditional approach. These systems often struggle with noise and variability in EEG data, making them less robust and reliable in real-world scenarios. Moreover, there is a notable research gap in integrating advanced machine learning (ML) and neural network techniques in the domain of brain-controlled keyboards. The literature reveals limited exploration of these approaches, especially in the context of voluntary hand movement-based assistive technology development, which is critical for enhancing functionality and accessibility. Addressing these issues requires adopting cutting-edge neural architectures capable of handling EEG data's intricacies while ensuring accuracy, scalability, and adaptability. Deep learning, with its proven ability to uncover complex patterns and adapt to diverse contexts, offers significant untapped potential to bridge these gaps and drive innovation in the field of BCI-based virtual keyboards.

The goal of this research is to leverage brain signals, specifically right and left voluntary hand movement signals, to control a virtual keyboard, making computer interfaces more inclusive and accessible. Our work introduces a novel EEG-based virtual keyboard system that employs a hybrid neural network architecture with Bidirectional GRU-Attention (BiGRU-Attention). Additionally, we perform comparative analysis with traditional ML models and advanced architectures, aimed at developing an optimized model for seamless integration into the application.

We demonstrate the practical potential of BCI-controlled virtual keyboards in improving the quality of life for individuals with neurodegenerative diseases and motor neuron disorders as shown in Figure 2. By interpreting brain signals from 19 electrodes, our system accurately simulates specific key presses, such as 'd' and 'l' or rest state. Such functionality highlights the system's capability as an efficient and reliable communication platform. Furthermore, the integration of hybrid neural networks ensures scalability and adaptability, enabling the system to evolve with advancements in EEG acquisition and signal processing technologies.

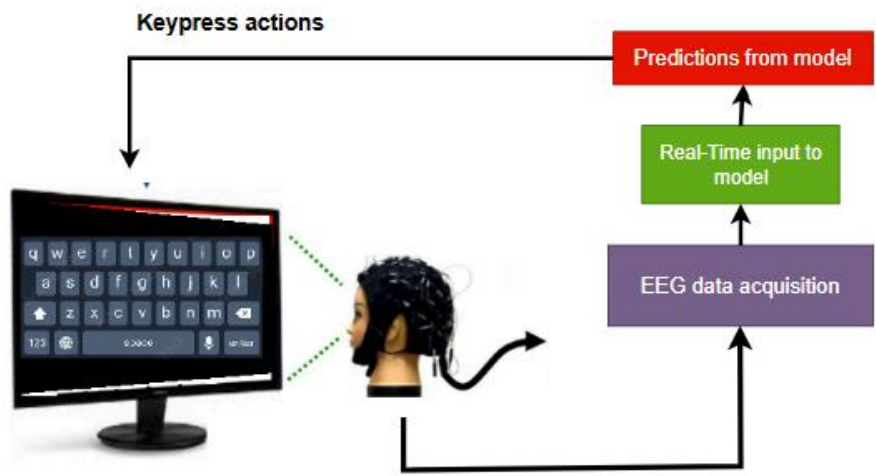

**Fig 2.** System practical application concept

The importance of this work lies in its ability to empower people with severe motor impairments, giving them greater independence and access to digital platforms. By bridging existing gaps in BCI development, this work advances human-computer interaction and assistive technology. A unique methodology for detecting right-left voluntary hand movements expands the capabilities of BCI-based assistive technologies and lays the foundation for innovative applications in assistive devices and beyond

## 2. Literature Review

Chambayil et al. [1] focus on a virtual keyboard system built using the LabVIEW platform, where eye blinks, traditionally considered artifacts, are repurposed as control signals. By analyzing the kurtosis coefficient and amplitude characteristics of eye blinks, the study demonstrates a novel approach to transforming these signals into actionable inputs for character selection, addressing key challenges in EEG artifact management.

Orhan et al. [2] developed the RSVP Keyboard, a BCI system that utilizes rapid serial visual presentation and language models to enable high-speed and accurate letter selection for individuals with LIS. By leveraging EEG-based ERP, the RSVP Keyboard aims to address the communication needs of this population where existing solutions fall short. Initial results demonstrate the potential for single-trial or few-trial accurate brain-based typing.

Hosni et al. [3] presented survey on the research on combining EEG and electro-oculography (EOG) signals in the development of hybrid brain-computer interface (hBCI) systems. Existing work on EOG-based human-computer interaction applications, particularly virtual keyboard interfaces, is organized and analyzed to identify the potential benefits of integrating EEG and EOG modalities. The authors propose a general architecture for a new EEG-EOG hBCI system that treats EOG traces as an additional input modality rather than artifacts to be removed, with the aim of inspiring the design of more practical and robust BCI systems.

Khong et al. [4] proposed a multi-player 3D video game that is controlled using both EEG-based inputs related to three different levels of attention, as well as conventional keyboard controls. Their experimental results demonstrate the feasibility of integrating brain signal-based inputs alongside traditional inputs to create a neurofeedback gaming system aimed at enhancing cognitive functions like attention through real-time EEG feedback

Junwei et al. [5] presented an EEG-based BCI system that can detect four mentally composed tasks using band power and radial basis function analysis of the EEG signals. Their proposed BCI system achieves an overall average classification accuracy of 92.50% for these four tasks, demonstrating the feasibility of using EEG-based control commands to operate intelligent systems such as a wheelchair for individuals with neurodegenerative conditions.

In order to help people with neuromuscular problems regain their mobility, Korovesis et al. concentrated on the construction of a brain-controlled mobile robot that uses alpha brainwaves for motion control [7]. Robotic motions in four directions—forward, backward, left, and right—are controlled by a synchronous, endogenous Electroencephalography (EEG) interface that reads eye blinks. A neural network is trained using features extracted from filtered EEG signals to provide precise robotic steering. With a high overall accuracy of 92.1%, experimental evaluations with 12 participants showed the system's efficacy and its potential to enhance the lives of those with severe disabilities.

Al-Turabi and Al-Junaid [8] presented the development of a complete BCI system that can control the movement of a smart wheelchair using non-invasive EEG brain signals. Their proposed BCI system employs three different ML algorithms - K-nearest neighbor, support vector machine, and artificial neural network - to classify the user's intention to move the wheelchair in different directions, with the support vector machine achieving the highest accuracy of 79.2%. The results demonstrate the feasibility of using EEG-based BCI for enabling independent mobility for individuals with physical disabilities.

Notturno et al. [11] examined the role of EEG as a potential marker for disease severity in amyotrophic lateral sclerosis (ALS). EEG was recorded in 15 ALS patients and 15 healthy controls while their eyes were closed. The researchers analyzed spectral band power in delta-theta, alpha, and beta frequency bands, as well as EEG microstate metrics, and correlated these features with clinical assessments of disability and disease progression. The results indicate that increased beta-band power in motor/frontal regions and altered microstate dynamics may be indicative of greater disease burden in ALS patients.

Saravanakumar and Reddy [12] presented the design of a novel hybrid speller/keyboard system that combines EOG and steady-state visual evoked potential (SSVEP) to enable selection of 36 targets, including letters, numbers, and special characters, divided into nine groups. The target group is selected using various eye movements (gaze, blinks, winks), and the specific target is then identified within the selected group using SSVEP, resulting in an average classification accuracy of 94.16% and an information transfer rate of 70.99 ± 9.95 bits/min, outperforming conventional EOG-based speller systems.

Wolpaw et al developed a BCI system for cursor control [13]. In this study, the 8–12 Hz mu rhythm recorded from the scalp over the central sulcus was used to investigate the creation of a novel communication and control system for people with significant motor deficiencies. To reach screen targets in about three seconds, subjects were trained to modify mu rhythm amplitudes, with greater amplitudes moving a cursor upward and lower amplitudes moving it downward. One significant advancement was the conversion of mu rhythm amplitudes into cursor movements using a distribution-based technique, which enabled accurate calibration and customized control parameters. Accurate 1-dimensional cursor control was attained by the subjects over weeks, demonstrating the potential of the mu rhythm as a communication tool for people with disabilities. Further performance improvement and 2-dimensional control are the goals of further training and parameterization improvements.

Mir et al. [14] explored the use of an EEG-based BCI, bionics, and Emotiv system integrated into a smartphone application as a potential solution to assist ALS patients. They aim to enable ALS patients, who experience progressive paralysis of the body, to regain a degree of control over their environment and facilitate basic communication with family members and caregivers through a computer interface. By leveraging EEG and related technologies, the research seeks to improve the quality of life for individuals affected by this debilitating neurodegenerative disease

Salih and Abdal [15] presented a BCI control tool that utilizes the Neurosky Mindwave headset to detect voluntary blinks and attention from the user's frontal lobe brainwaves. They aim to provide an alternative computer input mechanism for physically disabled individuals, exploring two virtual keyboard designs that enabled participants to achieve typing speeds of 1.55-1.8 words per minute, a promising result compared to previous BCI studies. The work demonstrates the potential of leveraging consumer-grade EEG devices to develop accessible input solutions for those with physical limitations.

Sharma et al. [18] proposed a BCI system that utilizes brain signal analysis to identify users and grant them autonomous control, enabling paralyzed or disabled individuals to operate a wheelchair by simply directing it using their brain activity. The prototype leverages the

Neurosky EEG sensor, which is more portable and easier to use compared to traditional EEG systems, to interface the user's brain with the wheelchair controls, providing an accessible mobility solution for those with physical limitations.

Palumbo et al. [19] presented systematic review that examines the state-of-the-art applications of EEG-based BCIs, particularly those utilizing motor-imagery data, for wheelchair control and mobility. The review provides a comprehensive overview of research conducted since 2010, analyzing the algorithms, feature extraction and selection techniques, classification methods, wheelchair components, and performance evaluation used in these BCI systems. The findings aim to shed light on the limitations of current biomedical instruments used to assist individuals with severe disabilities and identify novel research directions in this field.

Lin et al. [20] presented a novel "triple RSVP" BCI speller system that addresses the limitations of gaze-dependence and space-dependence in current matrix-based BCI spellers. The proposed system achieved an average online accuracy of 0.790 and an average online information transfer rate (ITR) of 20.259 bit/min, with a spelling speed of 10 seconds per character using a compact 90x195 pixel stimulus presentation interface, making it suitable for integration into mobile smart devices like smartphones and smartwatches.

Meng et al. [21] found that a group of 13 human participants were able to voluntarily modulate their brain activity to effectively control a robotic arm with high accuracy in reaching and grasping tasks, using a combination of two sequential low-dimensional control signals. The subjects were able to gain proficiency in controlling the robotic arm through brain rhythm modulation within just a few training sessions, and they maintained this ability over multiple months, demonstrating the feasibility of operating prosthetic limbs using non-invasive BCI technology.

Paneru and Paneru [22] presented a system that leverages computer vision techniques, including a neural network model trained on the ibug 300-W dataset, to detect keypress events and blinks in real-time, enabling communication for individuals with ALS and other motor function deficits. The system demonstrates effective blink recognition, user-triggered actions via Flask, and timely caregiver alerts through WhatsApp, showcasing its potential as an assistive technology to enhance the quality of life and communication for those suffering from neurodegenerative disorders. By providing practical solutions, the research advances healthcare technology and improves the standard of living for people with such conditions.

Dev et al. [23] presented the development of an EEG-based brain-controlled wheelchair using a BCI system and the NeuroSky MindWave EEG headset. The wheelchair's movement is regulated by the fluctuating attention levels of quadriplegic patients, who can also turn the device on and off through double eye blinks, enabling independent mobility for those with severe motor disabilities. The wheelchair's design incorporates a graphics-based fuzzy interface to help patients adjust their concentration levels as needed.

Intisar et al. [24] presents the design and development of a robotic vision system controlled through an interactive GUI application, aimed at enabling novice users to operate the system with minimal training. The application allows users to specify the object they want the robotic arm to pick up and place by filtering based on color, shape, and size, utilizing computer vision algorithms to determine the object's centroid coordinates, which are then used to control the arm's joint movements via a microcontroller. The goal is to create an automated robotic system that can be easily operated by users without extensive technical expertise.

Corley et al. [26] described the development of a virtual keyboard implemented using a BCI that interacts with the Emotiv EPOC neural headset, providing an alternative input device for individuals with motor disabilities who face challenges with traditional input methods. The authors summarize the advantages of a BCI-based virtual keyboard and present the design and implementation details, as well as the results of a preliminary study that identified areas for improving the effectiveness of the virtual keyboard system.

Reddy et al. [27] developed a BCI-based virtual keyboard with 36 keys designed according to the QWERTY layout, allowing users to control the keyboard using their brain's bioelectrical signals. Through a two-module approach involving EEG signal processing and hardware integration, the system achieved an average accuracy of 89.7%, a typing speed of 6.4 characters per minute, and proved effective in assisting individuals with neuromuscular disorders like paralysis, showcasing the potential of BCI technology to enhance communication and accessibility for those with physical disabilities.

Naseeb et al. [28] presents the development of a BCI-based virtual keyboard with 36 keys designed according to the QWERTY layout, allowing users to control the keyboard using their brain's bioelectrical signals. The system, composed of software and hardware modules, processes the user's EEG signals to detect RGB colors through an asynchronous mechanism, and experimental results showed an average accuracy of 89.7%, a typing speed of 6.4 characters per minute, and an average spelling completion time of 2.3 minutes, demonstrating the potential of BCI technology to assist individuals with neuromuscular disorders.

Scherer et al. [29] proposed a BCI-based virtual keyboard (VK) system that is asynchronously controlled by three classes of motor imagery and driven by a spontaneous electroencephalogram. The initial results from three able-bodied participants operating the VK showed that two of them were successful, demonstrating an improvement in the spelling rate up to 3.38 letters per minute on average, highlighting the potential of this approach to enhance brain-computer communication and develop more powerful applications.

Rusanu et al. [30] presents a novel LabVIEW-based algorithm for developing a BCI virtual keyboard controlled by the strength of eye blinks, aimed at assisting patients with neuromotor disabilities such as LIS or ALS. The virtual keyboard offers features like voluntary eye blink detection and counting, switching and selecting commands, highlighting concurrent actions, and enabling cancel, delete, and space commands, utilizing a "divide and conquer" approach to allow users to navigate through rows, half-rows, and individual keys to input characters.

Sharma et al. [31] proposed a transformer-based deep learning neural network architecture for motor imagery (MI) signal classification in EEG-based BCI applications, which aims to address the challenges of non-stationarity and long-term dependencies inherent in MI-EEG data. The proposed transformer-based model achieves superior performance compared to existing state-of-the-art methods, reaching classification accuracies of 99.7% on binary-class datasets and 84% on multi-class datasets, outperforming traditional LSTM-based approaches in capturing the long-term temporal patterns in MI-EEG signals.

Zhao et al. [33] introduced a Convolutional Transformer Network (CTNet) architecture for classifying EEG-based MI signals in BCI applications. The proposed CTNet model combines a convolutional module for extracting local and spatial features from EEG time series, followed by a Transformer encoder module leveraging multi-head attention to capture global dependencies, achieving remarkable decoding accuracies that outperform state-of-the-art methods in both subject-specific and challenging cross-subject evaluations on benchmark BCI datasets.

There is a notable research gap in developing EEG-based keyboards, with ML techniques being largely underexplored in the context of neuro-assistive keyboard systems. To address this, we propose the first EEG-based virtual keyboard utilizing the deep learning, capable of simulating keypresses using EEG data from new users. While previous research, such as [28], applied SVM for keyboard development, it remains the sole study employing ML in this context developed according to "QWERTY" standards which had 36 keys in total. Additionally, to the best of our knowledge, our work is the first to leverage right-left hand voluntary movement for EEG-based keyboard. Using signals from 19 electrodes, our deep learning model predicts brain activity to simulate keypresses for the 'd' and 'l' buttons.

## 3. Methodology

### 3.1 Data Acquisition

The process begins with data acquisition as shown in Figure 3. The first step involves obtaining an EEG dataset from the FreeForm paradigm in *Nature*, which captures neural activity associated with voluntary motor movements preceding key presses [6]. This dataset includes EEG signals, key press events, and associated labels, where "0" represents a blank screen (rest state), "1" indicates a "d" key press, and "2" denotes an "l" key press. The subjects interacted with an GUI that allowed for FreeForm interaction, focusing on a fixation point and freely pressing the "d" or "l" keys with their left or right hand at random intervals. The neural activity changes that preceded key presses were captured by the EEG signals, which served as the foundation for further analysis.

For data acquisition in the experiment conducted [6], Participants were first shown action signals that represented one of the mental imagery's that was going to be used. For the duration that the action signal was active, participants used the images once. EEG-1200 hardware captured the EEG signal associated with the implemented images, which was then stored using Neurofax recording software. The obtained EEG data were saved and exported as an ASCII file for additional processing following the experiment. Matlab was used to analyze the ASCII data file.

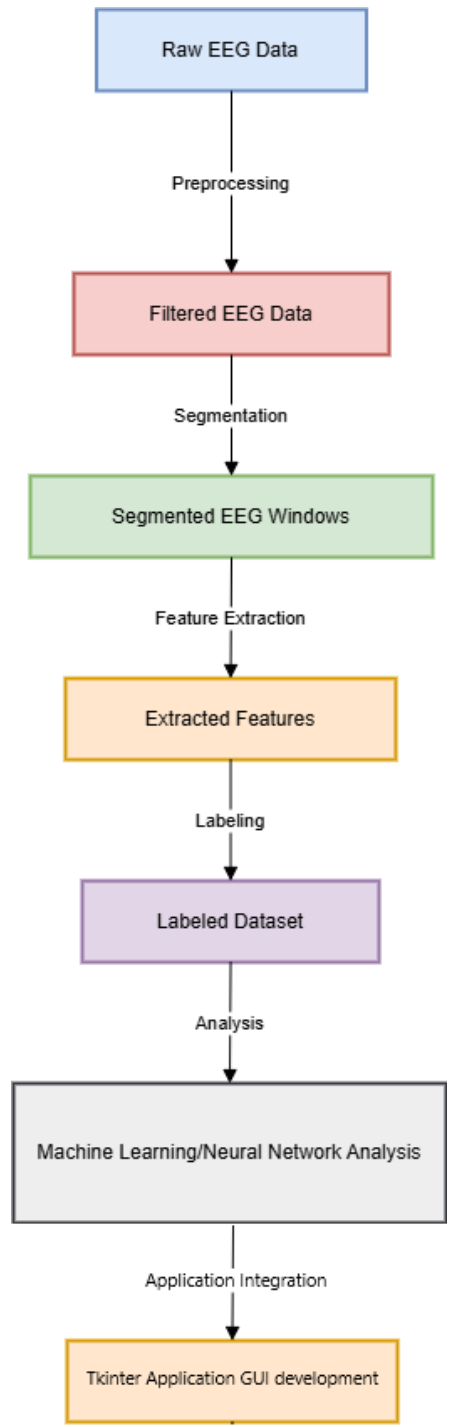

**Fig 3.** Proposed workflow

Figure 3 shows the workflow in which preprocessing and filtering have been carried out on the data obtained from [6], eliminating the need for extensive preprocessing.. The dataset consists of three recording sessions obtained from the FreeForm experiment. Table 1 summarizes the datasets and corresponding keystroke events. FreeForm—the identification of voluntary left-and right-hand movements—is the recording session paradigm. There are two states of mental imagery (2St). The total number of user keystroke events during the activity that corresponds to the stimuli or event during keypress or resting state is referred to as the onsets in the EEG recording. A crucial component of the classification problem is the onset information that was acquired.

Table 1. Onsets in the EEG dataset

| S.no. | Dataset | No. of Keystroke events |
|---|---|---|
| 1. | FREEFORMSubjectC1512082StLRHand | 688 |
| 2. | FREEFORMSubjectB1511112StLRHand.mat | 739 |
| 3. | FREEFORMSubjectC1512102StLRHand | 700 |

Each record in the dataset is distinguished by a unique alphanumeric identifier referred to as "id." This identifier serves as a key element for record tracking and management. The "nS" parameter denotes the number of EEG data samples contained within each record, providing insight into the temporal dimension of the recorded neural signals. The "sampFreq" parameter specifies the sampling frequency of the EEG data, representing the rate at which data points are collected per unit of time. The "marker" field encapsulates the GUI interaction record of the recording session, offering contextual information about user actions during the EEG data acquisition [6]. Finally, the "data" field encapsulates the raw EEG data for the respective recording session, serving as the primary source for subsequent preprocessing, feature extraction, and model training in the context of the BCI-Based Virtual Keyboard.

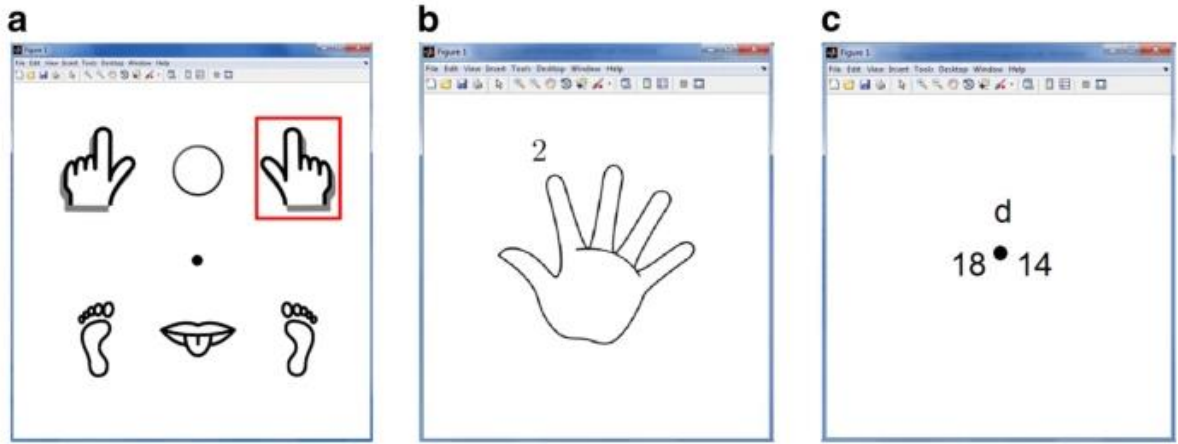

**Fig 4.** FreeForm-interaction GUI screen [6]

A crucial part of the BCI system was the easy GUI (Graphical User Interface), which enabled voluntary hand movement tasks based on mental imagery. A stimulus action signal, represented by a red rectangle indicating one of the icons representing the mental imagery to be used, was presented to the participant by the GUI at the start of each trial. During the one-second action signal display, participants were told to perform the specified mental imagery. The trial then ended with a varied pause of 1.5–2.5 seconds, resulting in an average trial duration of roughly 3 seconds. The GUI as shown in Figure 4 presented randomly selected mental imagery exercises to participants, guiding them through around 300 trials throughout each 15-minute engagement session. Continuous EEG signal recording was made possible by this controlled interaction, guaranteeing uniform task presentation and consistent timing throughout the three 15-minute parts of each 50–55-minute recording session [6].

With their hands lying on a keyboard and their left or right hand voluntarily pushing the "d" or "l" keys at random intervals, participants concentrated on a fixation point. EEG waves were analyzed using the key press times as reference points. Participants were encouraged to hit the left and right keys roughly equally by the GUI, which logged the total number of left and right key presses and showed the last key pushed. In this self-paced paradigm, changes in neural activity associated with motor planning and execution were seen in the EEG data just prior to the key presses. We note that choosing of 'd' and 'l' keypress in the GUI keypress is a random choice and does not have a depth association with any scientific concepts. The workflow is depicted in Figure 3.

## 3.2 Data Preprocessing and Feature Extraction

Following acquisition, data preprocessing is carried out. The EEG data is segmented into trials aligned with the key press events. Event plots with markers are used to identify and visualize these events, ensuring precise alignment. The dataset markers are tabulated in Table 2 which are actual classes for each data array. ERP plots are then utilized to observe and analyse neural activity changes before key presses. This step ensures that the data is well-prepared for feature extraction and model training.

**Table 2**: Markers in the dataset

| Marker | Event Activity |
| --- | --- |
| 0 | rest |
| 1 | 'd' |
| 2 | 'l' |

The feature engineering stage extracts relevant features from the pre-processed EEG data. Time-domain and frequency-domain features are considered to capture meaningful patterns in the neural signals. Notably, the EEG signal did not require additional special filtering, as it was pre-processed using hardware filters during data acquisition. The preprocessing and segmentation phenomenon can be seen in Figures 5 (a) and 5(b).

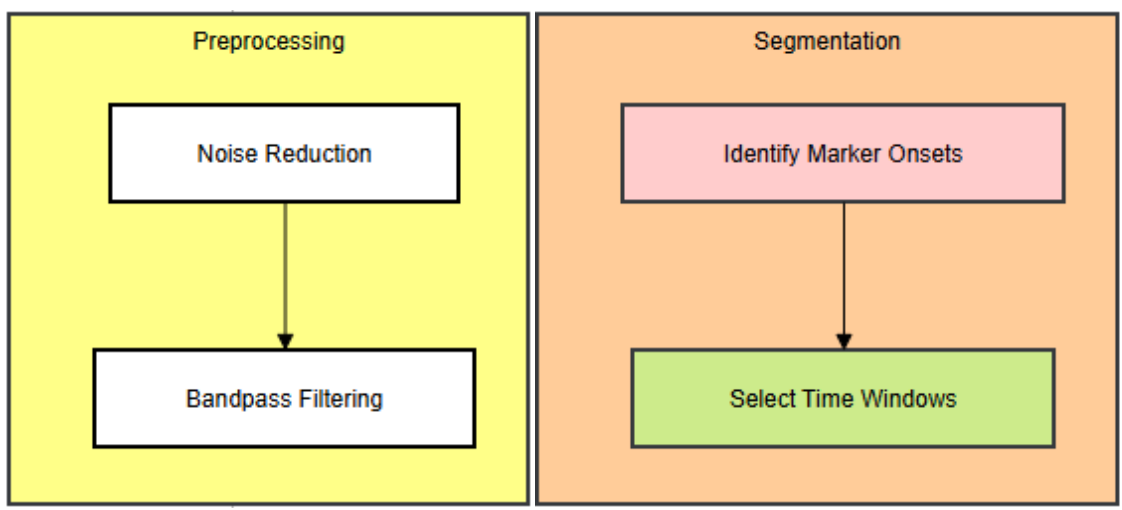

**Fig 5 (a).** Dataset segmentation phenomenon   **Fig 5(b).** Data Segmentation phenomenon

The Neurofax program applied a band-pass filter of 0.53–70 Hz at a 200 Hz sampling rate for most recordings, while a broader band-pass range of 0.53–100 Hz was used for signals sampled at 1000 Hz. Additionally, a 50 Hz notch filter in the EEG-1200 hardware reduced electrical grid interference. The dataset was refined by segmenting it into data from 19 electrodes, removing two channels irrelevant to brain activity-based control. ERP-based segmentation further divided the data into arrays, resulting in 19 × 3800 arrays. MATLAB file paths containing EEG data were iterated using a custom script, loading and processing data when the required information and labels were present.

Next, feature extraction and data saving involve extracting EEG features for each class onset identified in the marker data. A window around each onset is defined to capture relevant EEG data, including the onset and its aftermath. Extracted features are organized into rows along with their corresponding class labels, which are saved into a consolidated dataset for further use. The dataset is then prepared for ML by splitting it into training and validation sets, enabling effective model training and evaluation.

### 3.2.1 Sampling Frequency and Movement Onset Windows

The frequency at which the analog EEG signal is digitalized or sampled per second is known as the sampling rate in EEG. It is measured in Hertz (Hz) and establishes the fidelity and resolution of the recorded EEG data.

Sampling Rate $(Ts) = (1 / Ts)$

Where, $Ts$ is the sampling time in seconds.

The beginning of discernible alterations in EEG signals that signify the commencement of a particular event or activity in the brain, such as a seizure, a reaction to a stimulus, or the start of muscular preparation. The EEG onset movement window, as used in BCI applications, is the precise period during which brain activity associated with the start of a movement (or motor intention) is recorded and examined. When developing systems that decipher movement intentions for use in assistive devices, rehabilitation, or prosthetic control, this window is essential. On the basis of such an onset window, the EEG data is extracted with the help of a sampling rate i.e. 200. These onset times allow for the selection of EEG data segments from the "data" array by applying specified time offsets into the pre- and post-action signal on-time periods. This creates an EEG data fragment linked to the participant's provided mental imagery. The marker onsets utilized helped in segmentation and time windows were selected for the feature extraction.

During preprocessing, trials are extracted from the 200 Hz EEG data and aligned with recorded key press events. Event markers are plotted to visualize and identify these occurrences. ERP plots and additional analysis techniques are utilized to enhance signal quality and distinguish neural patterns. If the ERP results are inconclusive, advanced filtering methods are explored to improve data quality further. Following preprocessing, a deep learning model is trained for classification.

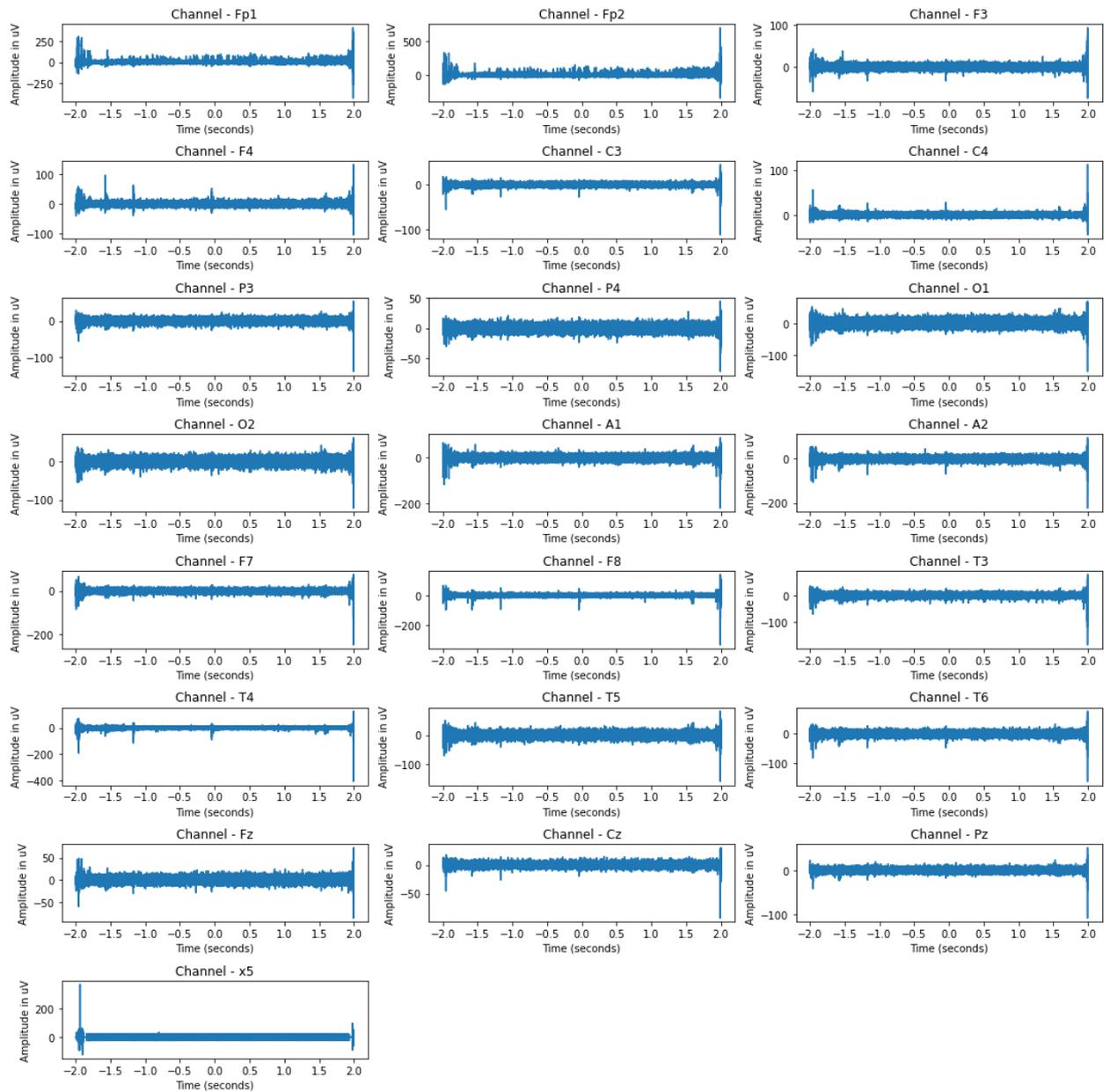

**Fig 6.** Visualization of the dataset

Figure 6 illustrates a 22-channel EEG recording, where each subplot represents electrical activity in the brain as measured by scalp electrodes. The channels are labeled according to the international 10-20 system, including positions such as Fp1, Fp2, F3, F4, C3, C4, P3, P4, O1, O2, A1, A2, F7, F8, T3, T4, T5, T6, Fz, Cz, and Pz, along with an additional channel labeled X5. Each channel corresponds to a specific brain region, and the subplots depict the amplitude in microvolts (μV) on the vertical axis and time in seconds on the horizontal axis, spanning a 2-second interval.

The EEG signals have been pre-filtered to remove high-frequency noise and artifacts, resulting in a smoother appearance with minimal distortion. The signals in each channel are stable and free of significant spikes or anomalies, indicating effective preprocessing. This likely involved bandpass filtering to isolate frequency bands associated with brain activity. By reducing the impact of external electrical noise and physiological distortions such as eye blinks or muscle movements, the preprocessing step enhanced the signal quality, ensuring that the data accurately reflects the brain's underlying activity. The redundant channels A1, A2, and X5 were removed concerning the 10-20 system, and the preprocessing enabled the authors to create models with greater consistency. The reason for omitting the 3 channels is described in Table 3.

**Table 3**: Omitted channels in the research study

| Channel | Reason for Omission |
|---|---|
| **A1** | The A1 channel, located at the left earlobe or mastoid in the 10-20 system, was omitted due to no good significance earlobes or mastoids, affecting its reliability in capturing brain activity. |
| **A2** | The A2 channel, corresponding to the right earlobe or mastoid in the 10-20 system, was excluded because it too doesn't have a great significance as placed on earlobes or mastoids, reducing its usefulness for ML model training. |
| **X5** | The X5 channel was not a standard EEG electrode in the 10-20 system and was primarily used for detecting movement or capturing EEG signals during specific conditions like sleep studies. It was not relevant to the primary focus of the research and thus omitted. |

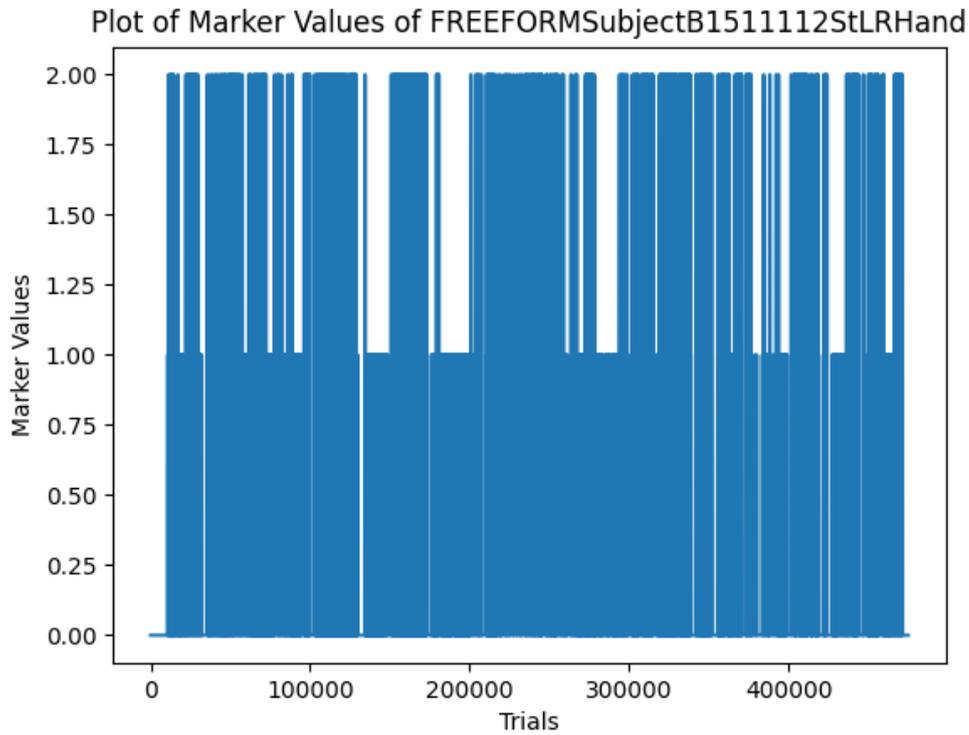

**Fig 7. Event plot of the data**

Figure 7. shows the plot of the event for freeform Subject B15111112StLRHand, which gives us important information about the activities and events during the recording session. "0," which denotes the resting state, is obtained when the "d" and "l" presses are made simultaneously. This allows us to observe that the volunteer participating in the recording session initially rests and other corresponding activities are initialized as the session progresses.

**3.2.2 ERP analysis**

We plotted the ERPs of 19 electrode data and analyzed the neural activity patterns, related to the key press. ERPs provide us average electrical activity of the brain in response to specific stimuli or events. The utilization of ERP plots aids in the identification of distinctive features and characteristics in the EEG data that correspond to motor planning and execution. These plots can reveal subtle changes in brain activity in the time domain, particularly in the moments immediately preceding the key presses. Analyzing ERPs helps in determining the temporal dynamics of the brain's response to motor intentions, which is crucial for feature extraction and model training.

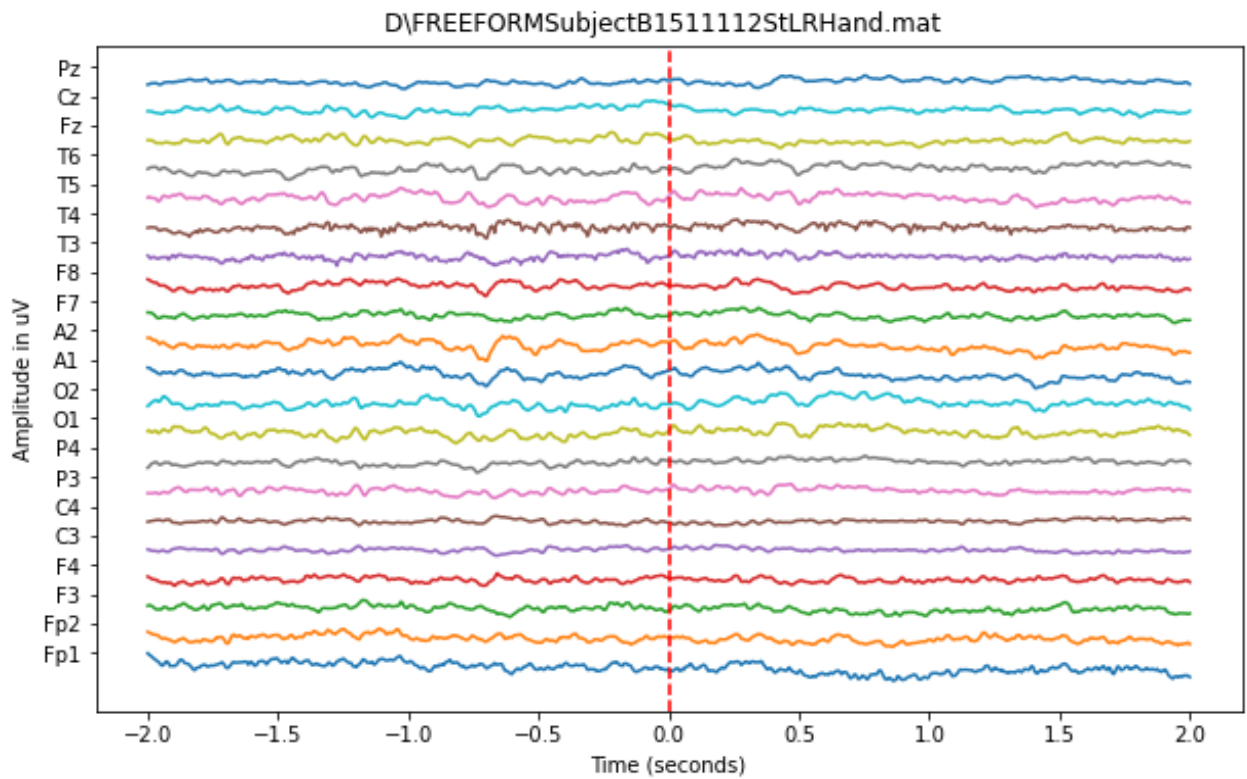

**Fig 8. ERP plot visualization**

The ERP plot, shown in Figure 8, shows the 21-channel EEG data plot concerning movement onset, which provide us with the behavior of the events in the recording session. The sampling frequency of 200 Hz is utilized to grab the samples and make analysis of EEG data. The ERP waveforms of a (FREEFORMSubjectB1511112StL) during a freeform task are displayed in the figure above. Plotting of the ERP waveforms for 19 electrodes on the subject's scalp is done with the ERP amplitude on the y-axis and the time axis on the x-axis.

The ERP provides information about the subject's brain activity during the freeform task. The peaks and valleys in the waveforms indicate various electrical activity patterns, and the various colors correspond to various scalp electrodes. For instance, the Pz electrode, which is located at the top of the head, exhibits a positive peak at approximately 300 milliseconds, which is commonly connected to the ERP's P300 component. It is believed that the P300 component reflects the subject's focus on the freeform task.

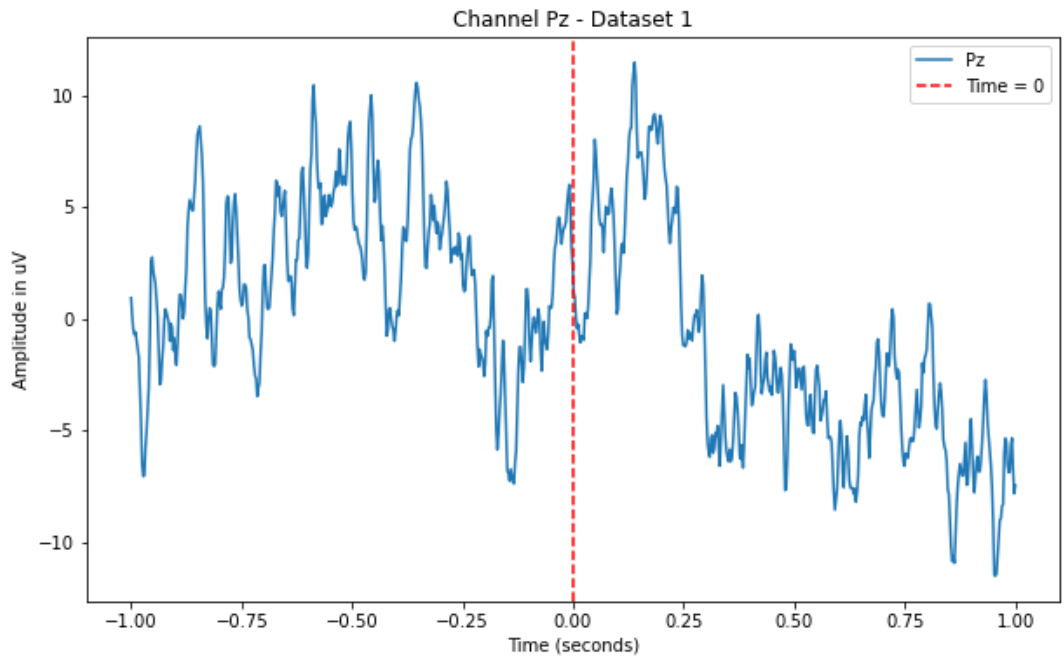

**Fig 9. ERP plot of channel Pz of dataset-1**

The Figure 9 shows an ERP waveform for a single channel (Pz) of a dataset. The ERP is a measure of the electrical activity of the brain in response to a specific event or stimulus. It is calculated by averaging the EEG data from multiple trials, which cancels out random noise and reveals the underlying brain response. The ERP waveform is typically characterized by a series of peaks and troughs, each of which is associated with a different stage of cognitive processing. For example, the P1 wave is thought to reflect the initial sensory processing of a stimulus, while the N1 wave is thought to reflect the attention to that stimulus. Later waves, such as the P300 and N400, are thought to reflect higher-level cognitive processes, such as decision-making and memory retrieval. The specific ERP waveform that is observed can vary as it is dependent on the type of stimulus or event that is presented. For example, the ERP waveform for a visual stimulus will be different from the ERP waveform for an auditory stimulus. The ERP waveform can also be affected by the individual's cognitive state, such as their attention level or fatigue.

The y-axis of the graph is in microvolts (µV), and the x-axis is in seconds. The graph shows a negative voltage deflection, or ERP, that peaks around 200 milliseconds after the stimulus. This ERP is called the P2 component. The P2 component is thought to reflect the brain's automatic processing of sensory stimuli. It is larger for attended stimuli than for unattended stimuli, and it is also sensitive to the complexity of the stimuli. The specific meaning of the P2 component is associated with the keypress events in the context of this project. For example, as the stimuli were words, the P2 component is larger for words that are attended to or that are unexpected. Overall, the image suggests that the participants in the experiment were paying attention to the stimuli and that their brains were processing them automatically. The ERP analysis visualizations for all channels are shown in Figure 10.

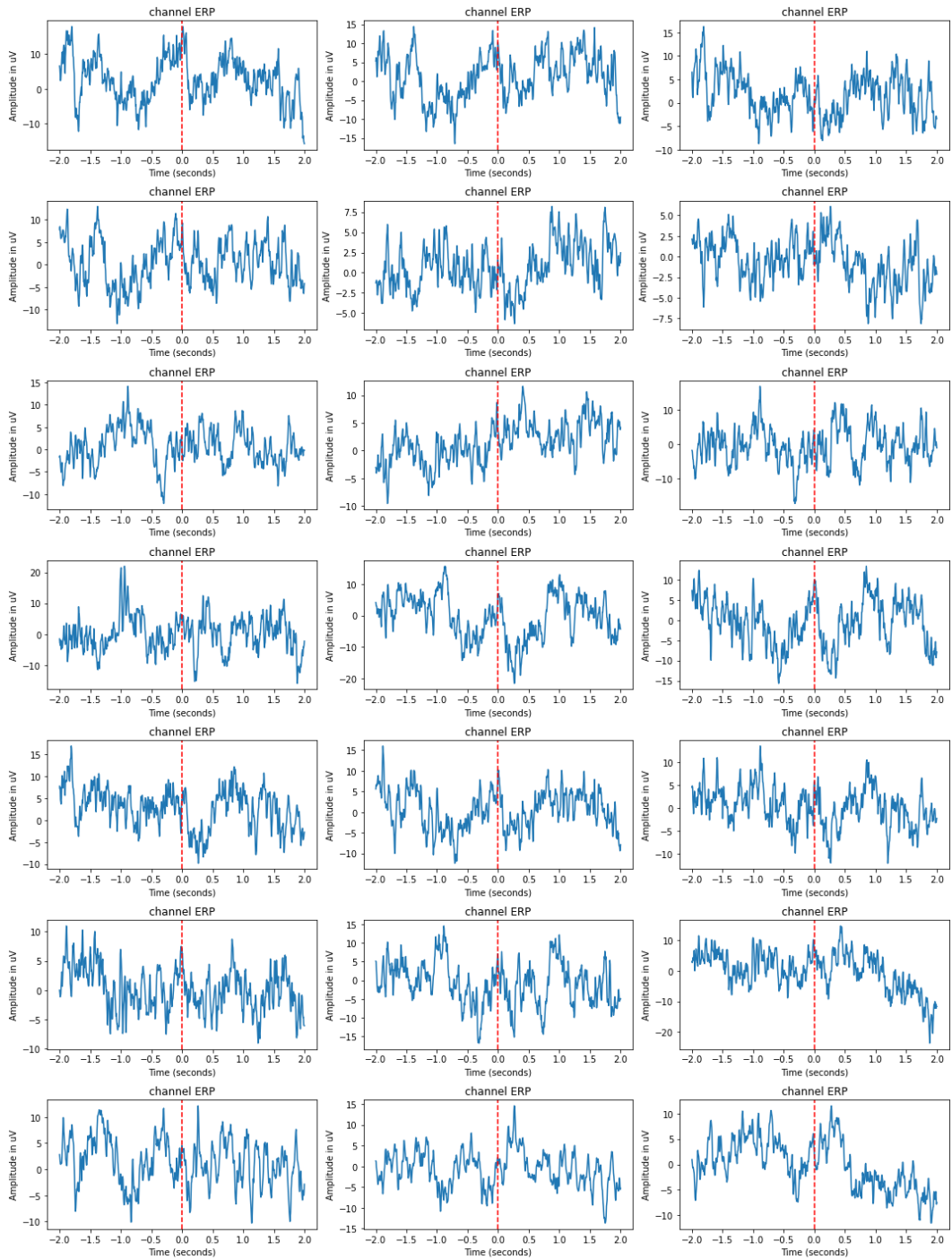

**Fig 10.** ERP plot visualization for all the channels

## 3.3 Machine Learning Models

### 3.3.1 SVM

SVM is a supervised ML algorithm widely employed for its effectiveness in handling high-dimensional datasets. SVM constructs a hyperplane or a set of hyperplanes in a high-dimensional space to separate data points into distinct classes, as illustrated in Figure 10. SVM is particularly effective in scenarios with more features than samples and in high-dimensional spaces. It supports various kernel functions, such as linear, polynomial, radial basis function (RBF), and sigmoid, to handle both linear and non-linear classification tasks. In our work, we used SVM with a linear kernel $K(x_i, x_j) = x_i^T x_j$, and $C = 0.001$, which provided computational efficiency and was well-suited for the dataset. Figure 11(a) shows the model architecture.

### 3.3.2 Gaussian Naïve Bayes (GNB)

GNB is a probabilistic supervised learning algorithm based on Bayes' theorem, widely used for classification tasks due to its computational efficiency and simplicity. Bayes' theorem is expressed as $P(C|X) = \frac{P(X|C)P(C)}{P(X)}$, where P(C|X) is the posterior probability of class C given features X, P(X|C) is the likelihood, P(C) is the prior, and P(X) is the evidence. GNB further assumes that continuous features follow a normal distribution, with the likelihood modeled $P(X|C) = \frac{1}{\sqrt{2\pi\sigma^2}} e^{-\frac{(X-\mu)^2}{2\sigma^2}}$, where μ and $\sigma^2$ are the mean and variance for each feature within a class. To ensure robustness, we applied 10-fold stratified cross-validation with shuffle, randomly splitting the data into 10 subsets for training and validation. This approach minimized bias and provided reliable classification performance, demonstrating the simplicity and effectiveness of GNB for the task. The architecture is shown in Figure 11(c).

### 3.3.3 Multilayer Perceptron (MLP)

MLP is a fundamental type of Artificial Neural Network (ANN) widely used in ML for tasks such as regression, image recognition, and sentiment analysis. An MLP consists of an input layer, one or more hidden layers, and an output layer, with each layer comprising interconnected nodes. The algorithm incorporates activation functions such as ReLU, sigmoid, or tanh to introduce non-linearity and capture complex relationships in data. For this work, the ReLU activation function was used, defined as $f(x) = \max(0, x)$, to ensure efficient gradient flow and mitigate the vanishing gradient problem. MLPs are trained using the backpropagation algorithm, which optimizes the loss function by adjusting weights via gradient descent. For a weight , the update rule is given by $w = w - \eta \frac{\partial L}{\partial w}$, where η is the learning rate and $\frac{\partial L}{\partial w}$ is the gradient of the loss L concerning w. In this work, an MLP was implemented with three hidden layers containing 256, 128, and 64 nodes, respectively, each employing ReLU activation and L2 regularization (λ=0.01) to prevent overfitting. Dropout layers with a rate of 0.2 were added after each hidden layer to further regularize the model. The network was compiled using the Adam optimizer, sparse categorical cross-entropy loss function, and accuracy as the evaluation metric. Training was performed over 100 epochs with a batch size of 128, using early stopping to monitor validation loss and prevent overfitting. The implementation was carried out using TensorFlow/Keras. The model architecture is given in Figure 11(b).

### 3.3.4 Categorical Boosting (CatBoost)

CatBoost is a powerful gradient-boosting algorithm designed to handle both categorical and numerical data efficiently. Unlike traditional methods, it eliminates the need for one-hot encoding by directly processing categorical features using a technique called ordered boosting, which reduces overfitting. CatBoost minimizes a loss function, such as Logloss for classification tasks, defined as $Logloss = -\frac{1}{N}\sum_{i=1}^{N}[y_i \log(p_i) + (1 - y_i) \log(1 - p_i)]$, where $y_i$ is the true label, $p_i$ is the predicted probability, and N is the number of samples. In our work, a CatBoost model with 1000 iterations, a depth of 6, and a learning rate of 0.1 was trained on EEG data, achieving high accuracy with minimal preprocessing, highlighting its robustness and computational efficiency. The Figure 11(d) shows CatBoost model architecture.

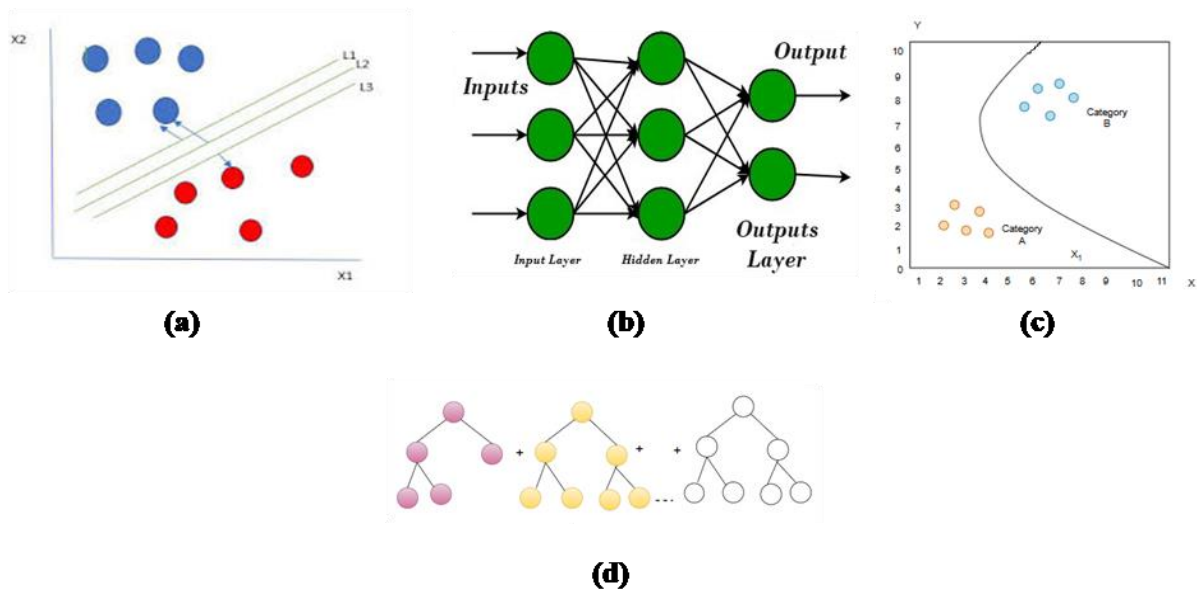
(a)      (b)      (c)

(d)

Fig 11: Schematic Diagram (a) SVM,(b) MLP,(c) NB ,(d) CATB

### 3.3.5 BiGRU-Attention Hybrid Model

The proposed BiGRU-Attention model integrates Bidirectional Gated Recurrent Units (BiGRU) with an attention mechanism to handle sequential data effectively. BiGRU-Attention model architecture is shown in Figure 11. Unlike traditional GRUs, BiGRUs process input sequences in both forward and backward directions, enabling the model to capture past and future temporal dependencies simultaneously. This enhances the model's ability to interpret complex sequential patterns in EEG data.

BiGRUs are designed to capture temporal dependencies using gating mechanisms, such as the update gate ($z_t$) and reset gate ($r_t$), which control the flow of information. The BiGRU hidden state ($h_t$) is computed as:

$$z_t = \sigma(W_z \cdot [h_{t-1}, x_t] + b_z)$$
$$r_t = \sigma(W_r \cdot [h_{t-1}, x_t] + b_r)$$
$$\widetilde{h}_t = \tanh(W_h \cdot [r_t \odot h_{t-1}, x_t] + b_h)$$
$$h_t = (1 - z_t) \odot h_{t-1} + z_t \odot \widetilde{h}_t$$
$$h_t^{\text{BiGRU}} = \left[h_t^{\text{forward}}, h_t^{\text{backward}}\right]$$

Where $h_t$ is the hidden state at time t, σ is the sigmoid activation, and ⊙ denotes element-wise multiplication. The final BiGRU hidden state $h_t^{\text{BiGRU}}$ is a concatenation of the forward $h_t^{\text{forward}}$ and backward $h_t^{\text{backward}}$ hidden states, allowing the model to capture temporal dependencies in both directions. This gating structure ensures efficient learning of sequential patterns while avoiding vanishing gradients.

The attention mechanism enhances the BiGRU by dynamically prioritizing the most relevant time steps. The attention weights ($\alpha_t$) are calculated as:

$$\text{score}_t = W \cdot h_t^{\text{BiGRU}} + b$$
$$\alpha_t = \frac{\exp(\text{score}_t)}{\sum_{t=1}^{T} \exp(\text{score}_t)}$$
$$c = \sum_{t=1}^{T} \alpha_t \cdot h_t^{\text{BiGRU}}$$

Where c is the context vector that aggregates the weighted contributions of all hidden states. The attention mechanism ensures the model focuses on critical temporal features, improving classification performance. The architecture, comprising a BiGRU layer, attention mechanism, and dense output layer with softmax activation, is trained using sparse categorical cross-entropy loss with L2 regularization to prevent overfitting.

$$L = -\frac{1}{N}\sum_{i=1}^{N}\log(p_{i,y_i})$$

This model demonstrates high efficacy in tasks such as EEG-based sequential data classification by combining the temporal modeling of BiGRUs with the dynamic feature selection of attention.

Our proposed model consists of a BiGRU layer containing 128 units to capture temporal dependencies in both forward and backward directions. An attention mechanism is added to dynamically compute weighted contributions of the hidden states, enabling the model to focus on the most relevant time steps. Dropout layers with a rate of 0.2 are included after the BiGRU and attention layers for regularization. The model was compiled using the Adam optimizer, sparse categorical cross-entropy loss function, and accuracy as the evaluation metric. Training is conducted with a batch size of 128, employing early stopping to monitor validation loss and prevent overfitting. The model architecture is shown in Figure 12.

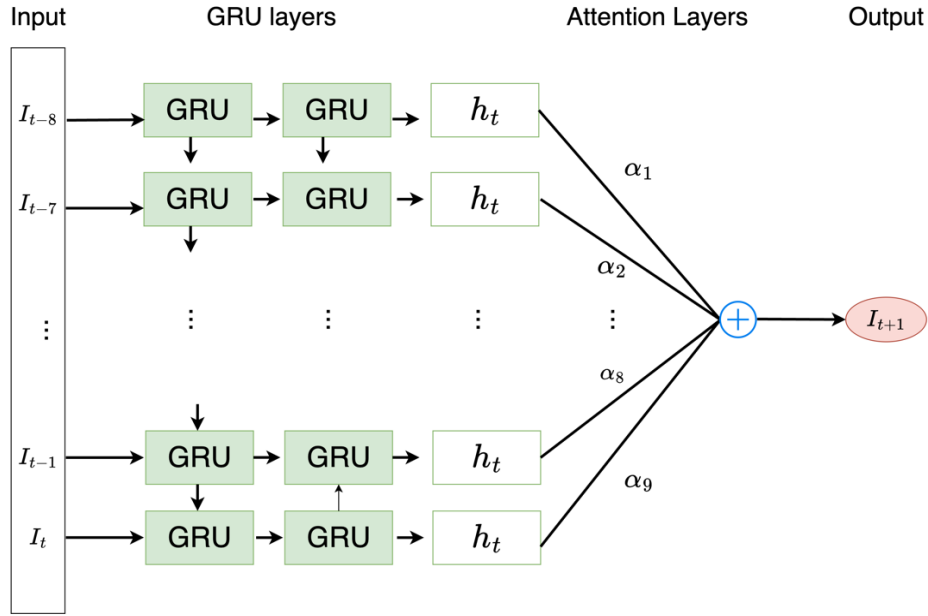

**Fig 12**. GRU-Attention hybrid model architecture

### 3.3.6 Baseline model development

In addition to these approaches, we employ a transformer-based deep learning neural network architecture [31], EEGNet [32], and a convolutional transformer network [33] as baselines. These baseline architectures are recognized as state-of-the-art for handling sequential data and are frequently applied to EEG datasets. Their strength lies in their ability to extract local spatial patterns using convolutional layers while simultaneously capturing long-range temporal dependencies through transformer layers. By incorporating these transformer-based methods, we enable a comprehensive evaluation of our models, benchmarking their performance against advanced architectures in EEG classification tasks. In addition to the transformer-based model, traditional models such as SVM, Naïve Bayes, CatBoost, and MLP are also used for comparative analysis.

### 3.4 Model Performance Evaluation Metrics

The performance should be assessed using statistical measures, classification metrics, hyperparameter tuning, and cross-validation techniques. The mean is a measure of central tendency calculated by dividing the sum of all values in a dataset by the total number of values. It provides a single representative value for the dataset but can be sensitive to outliers. Standard deviation measures the spread or dispersion of values around the mean. A high standard deviation indicates that the values are widely spread, while a low standard deviation suggests they are clustered close to the mean.

A confusion matrix is a table used to assess a classification model's performance. It displays an overview of the model's predictions in comparison to the actual factual values. Each row represents the actual class in the matrix, whereas each column represents the anticipated class. True Positive (TP), False Positive (FP), True Negative (TN), and False Negative (FN) are the four quadrants of the confusion matrix.

**(a)** This statistic assesses a model's effectiveness in binary classification tasks by combining precision and recall. It is calculated using the harmonic mean of recall and precision. When there is an imbalance between the classes in the dataset, the F1-score is helpful since it balances false positives and false negatives, as given by equation 1.

$$\text{F1 Score} = 2 \times \frac{Precision * Recall}{Precision + Recall} \qquad (2)$$

**(b)** A model's capacity to accurately identify every pertinent class instance is gauged by the recall, also referred to as sensitivity or true positive rate. The ratio of true positives to the total of false negatives and true positives is used to compute it. A high recall suggests that the model reduces false negatives well as given by equation 2.

$$\text{Recall} = \frac{TP}{(FN+TP)} \qquad (3)$$

**(c)** Precision is the capacity of a model to accurately identify, out of all the examples it has identified as belonging to a class, only the relevant instances of that class. The ratio of true positives to the total of true positives and false positives is used to compute it. A high precision means that the model reduces false positives as well as possible and is given by equation 3.

$$\text{Precision} = \frac{TP}{(FP+TP)} \tag{4}$$

**(d)** Accuracy is the proportion of cases that were accurately predicted in all instances. It is given by equation 4.

$$\text{Accuracy} = \frac{TP+TN}{(FP+FN+TP+TP)} \tag{5}$$

**(e)** A classification report summarizes a model's performance across multiple classifications. Typically, metrics like support for each class, accuracy, recall, and F1-score are presented. The report provides insights into the model's performance for each class, which can help identify areas where the model may need to be developed.

**3.5 Hyperparameter Tuning and Cross-validation**

Determining an ML algorithm's ideal hyperparameters is known as hyperparameter tuning. Hyperparameters are predetermined and cannot be discovered by data-driven analysis during the training phase. The tuning hyperparameters include the regularization parameter in linear models, the learning rate in neural networks, and the depth of decision trees. Hyperparameter tuning entails utilizing methods like grid search, random search, or more sophisticated optimization algorithms like Bayesian optimization to search over a predetermined range of hyperparameters. The objective is to determine which set of hyperparameters gives the model the highest performance, either through cross-validation or on a validation set.

A method for evaluating how effectively a prediction model generalizes to a separate dataset is cross-validation. The original dataset is split into k equal-sized subsets, or folds, for k-fold stratified cross-validation. On k-1 folds, the model is trained, and on the remaining fold, it is validated. This procedure is carried out k times, with a distinct fold serving as the validation set each time. The performance measures are then averaged over the k iterations to approximate the total performance.

**3.5  GUI development and integration**

The GUI is built using Python's Tkinter library, a standard and widely used tool for developing desktop applications. Tkinter simplifies the creation of interactive and user-friendly interfaces through widgets like buttons, labels, text boxes, and menus. The focus of this work is on classifying EEG data to simulate right- and left-hand voluntary movements using an ML model. The dataset, obtained from a Nature Journal publication, provides the basis for training and testing. Since no hardware components are involved, the testing data is represented as numpy arrays. Each dataset is saved as a .npy file, which the GUI dynamically loads during runtime. The GUI serves as an intuitive platform for feeding testing data into the ML model. It processes the EEG input, predicts the corresponding hand movement, and simulates the associated keypress events. By presenting the classification results clearly and interactively, the interface effectively demonstrates the model's capabilities and ensures a seamless user experience. This integration bridges the gap between EEG data analysis and practical application, offering a streamlined approach to visualizing model performance

4. **Results and Discussions**

After successive data preprocessing and feature engineering of 19-electrode EEG signals, the datasets were structured into rows and columns. The data was split into an 80-20 ratio, where 80% was used for training and 20% for testing. Additionally, 20% of the training set was reserved for validation to optimize model performance. Subject C-I and Subject C-II datasets were used for training, while the Subject B dataset served for validation.

The EEG signals, acquired at a 200 Hz sampling rate via the Neurofax program, underwent hardware-based pre-filtering. A band-pass filter (0.53–70 Hz) and a 50 Hz notch filter, provided by EEG-1200 hardware, reduced electrical grid interference [6]. These filters were applied during acquisition, ensuring consistency across all publicly available records. Preprocessing and feature extraction benefited from the hardware filtering, allowing the SVM model to achieve high performance. The optimal hyperparameters are summarized in Table 4.

**Table 4:** Best hyperparameter for all applied model

| Hyperparameter | SVM | NB | CATB | MLP | CTNet | Transformer-based | EEGNet | BiGRU-Attention |
|---|---|---|---|---|---|---|---|---|
| Penalty | - | - | - | - | L2 (0.03) | L2 (0.02) | L2 (0.03) | L2 (0.01) |
| Solver | - | - | - | - | Stochastic Gradient Descent | Adam | Adam | Adam |
| Max Iterations (max_iter) | - | - | 1000 | - | 500 | 50 | 125 | 100 |
| Multi-Class Handling | - | - | - | - | Softmax | Softmax | Softmax | Softmax |
| Random State | - | 42 | 42 | - | 42 | 42 | 42 | 42 |
| Shuffle | - | True | - | - | True (train-test split with stratify) | True (train-test split with stratify) | True (train-test split with stratify) | True (train-test split with stratify) |
| Standard Scaler | - | StandardScaler() | - | - | StandardScaler | StandardScaler | Applied channel-wise | StandardScaler |

| | | | | | | | for time points | |
|---|---|---|---|---|---|---|---|---|
| **Model** | - | GNB | - | - | CTNet (Convolutional + Transformer-based network) | Transformer-based | EEGNet | BiGRU (128) + Attention |
| **Depth** | - | - | 6 | - | - | 4 Transformer block | - | 1-layer BiGRU + Attention |
| **Learning Rate** | - | - | 0.1 | - | 0.0001 | Cosine decay scheduler (initial rate of 1e-3 and alpha 1e-6) | 0.001 | 0.001 |
| **Loss Function** | - | - | MultiClass | - | categorical_crossentropy | sparse_categorical_crossentropy | sparse_categorical_crossentropy | sparse_categorical_crossentropy. |
| **Eval Metric** | - | - | Accuracy | - | Accuracy | Accuracy | Accuracy | Accuracy |

| Dropout | - | - | - | - | 0.6 | 0.6 | 0.5 | 0.2 |
|---|---|---|---|---|---|---|---|---|
| Batch Size | - | - | - | - | 32 | 32 | 128 | 128 |
| Epochs | - | - | - | - | 500 (with early stopping) | 50 (with early stopping) | 125 (with early stopping) | 100 (with early stopping) |
| Validation Split | - | - | - | - | 0.2 | 0.2 | 0.2 | 0.2 |

The proposed model BiGRU-Attention shows a great performance with 90% on the testing set. The results of the 10-fold cross-validation for this hybrid model, presented in Table 5, demonstrate consistent and robust performance across different subsets of the data. The test accuracies for each fold range from 0.87 to 0.93, indicating that the model maintains high accuracy on various splits of the dataset. The mean test accuracy across all folds is 91%, demonstrating that the model performs consistently well overall. The standard deviation of the test accuracies is 0.02, which is low and indicates that the model's performance is stable and does not vary significantly across different folds. These results suggest that the BiGRU-Attention hybrid model is reliable and effective for the classification task, with minimal variability in performance across different data subsets.

## 4.1 Model Performance Analysis

### 4.1.1 SVM

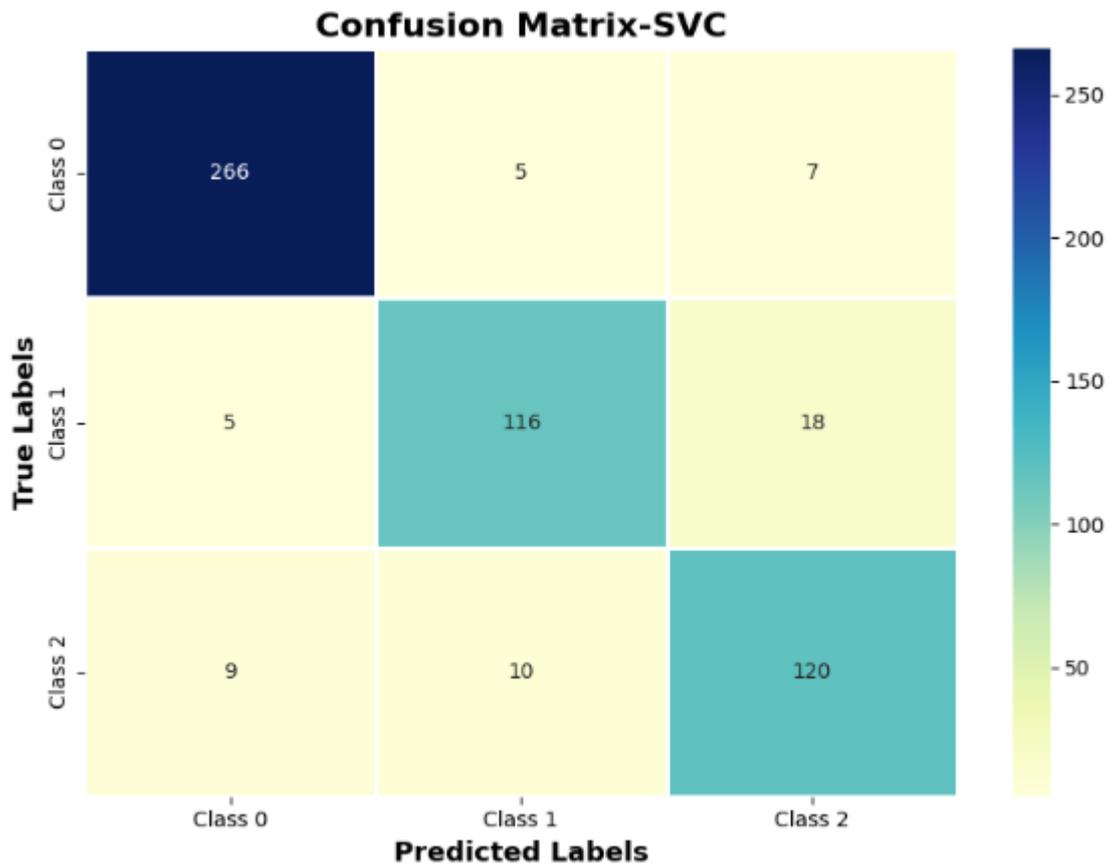

**Fig 13.** Confusion Matrix plot for SVC

The SVM model achieved 98% on training and similarly, 89% on testing. The confusion matrix plot for the SVC model is shown in Figure 13. The model was trained using a linear kernel, 10 splits and a hyperparameter value of $C=0.001$. The confusion matrix plot as shown in the figure shows very good results with 266 instances out of 278 total predicted correctly for class '0' that is the rest state, similarly, out of a total of 139 instances 116 were predicted correctly as class '1' is keypress 'd' action and finally, out of total 139 instances 120 instances were predicted correctly as class '2' which is keypress 'l' action. This indicates very high performance for the SVC classifier.

The optimal hyperparameter for the SVM model that was found via successive fine-tuning is shown in Table 4. Overall, the model appears to be operating effectively and stable across various subsets of the EEG signal data, as seen by its mean accuracy of 0.90 and low standard deviation of 0.016. This shows that the model can generalize well to fresh, untested samples and has picked up useful patterns from the data.

### 4.1.2 Gaussian Naïve Bayes (GNB)

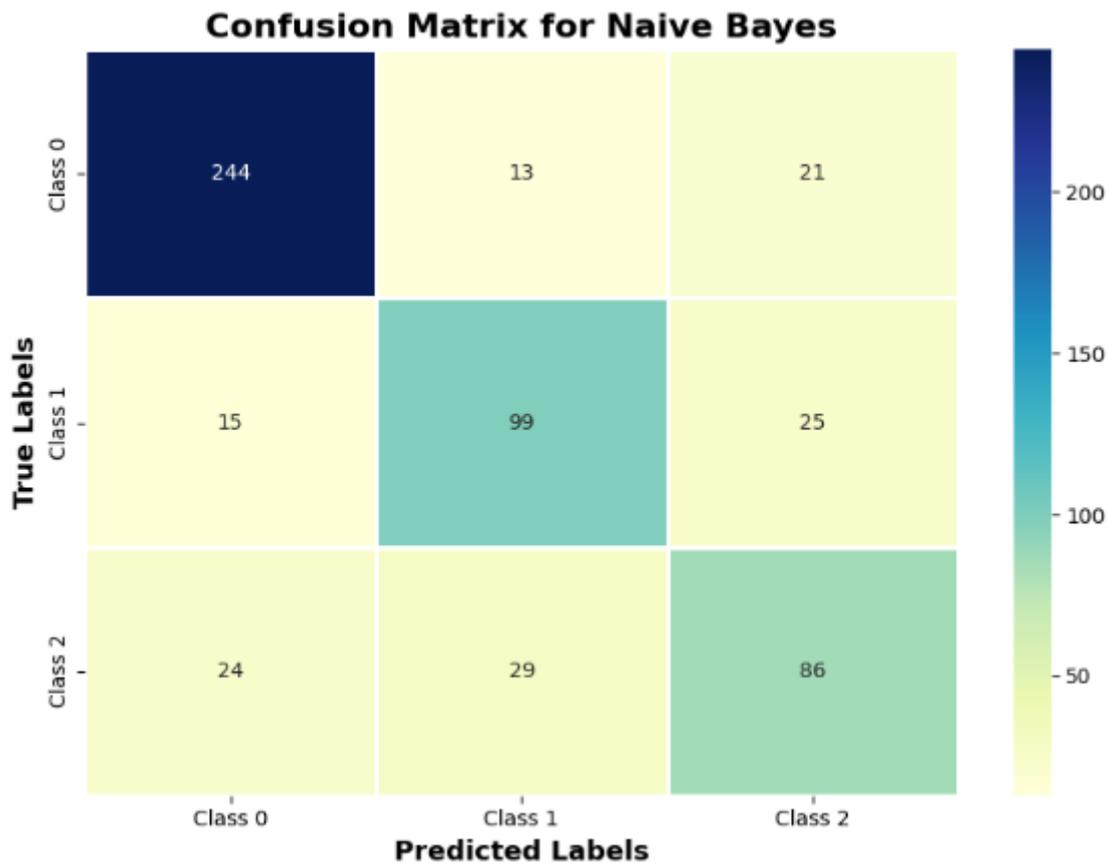

Figure 14: Confusion matrix plot for Naïve Bayes model

The confusion matrix plot, as shown in Figure 14, demonstrates the performance of the GNB model in classifying EEG-based events. For class '0', 244 instances were correctly predicted out of the total, showing strong performance in this category. For class '1', 99 instances were correctly predicted out of 139, with a moderate level of misclassifications. Lastly, for class '2', 86 instances were predicted correctly, though some misclassifications occurred, particularly into class '0' and class '1'. Despite these misclassifications, the overall results indicate the model's reasonable capability in classifying EEG-based events. The hyperparameters used for this GNB model are outlined in Table 4. The mean testing accuracy with stratified cross-validation is 0.79 with standard deviation of 0.0181 as shown in Table 5.

### 4.1.3 Catboost:

The CatBoost model obtained an accuracy of 88% on testing and 100% on training set and further evaluation with stratified 10-fold cross-validation following results as shown in Table 5.

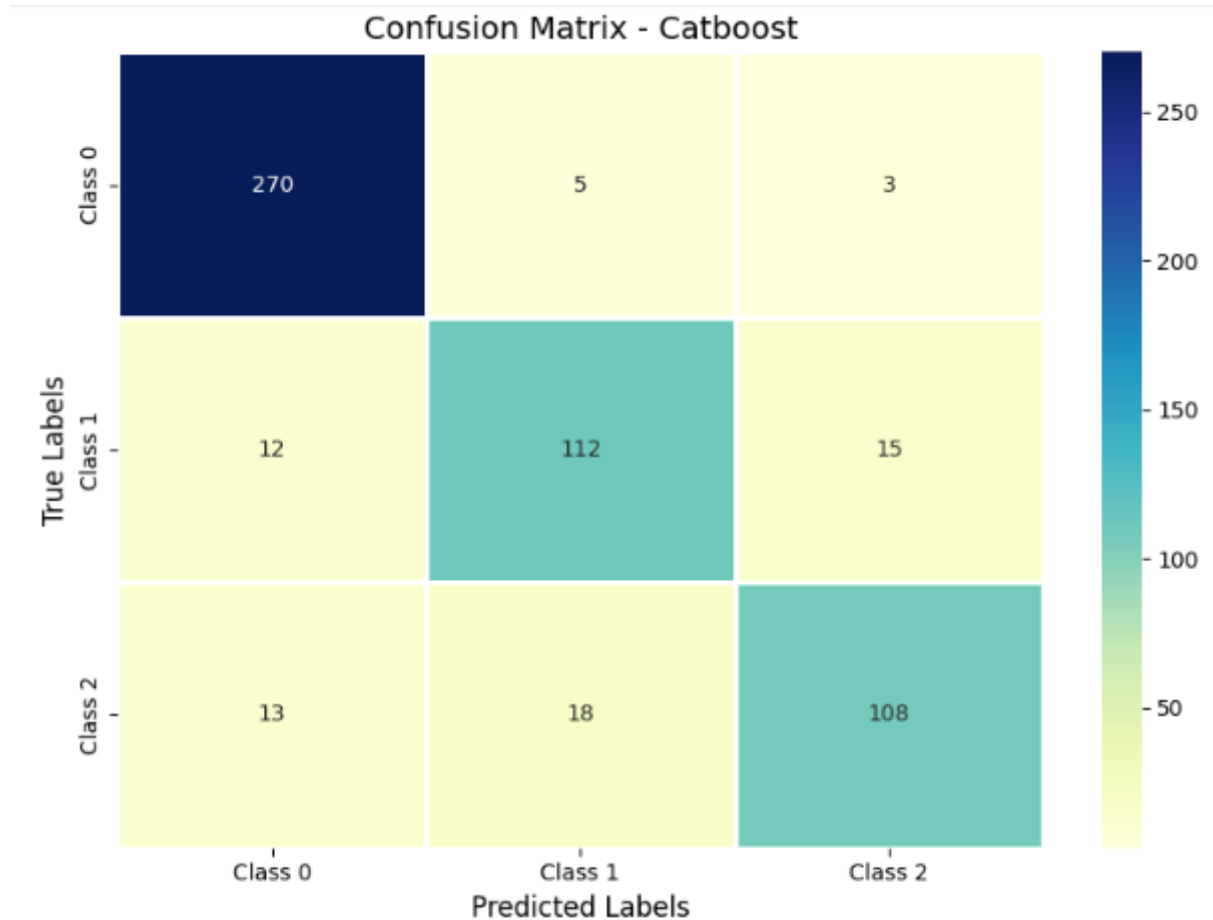

**Fig 15.** Confusion matrix plot for CatBoost model performance

The confusion matrix plot as shown in Figure 15, shows very good results with 262 instances out of 278 total predicted correctly for class '0' which is the rest state, similarly, out of total of 139 instances 120 were predicted correctly as class '1', which is keypress 'd' action. Finally, out of total 139 instances 120 were predicted correctly as class '2' which is keypress 'l' action. This shows the great performance and reliability of utilizing CatBoost on classifying the EEG-based events. The best hyperparameters used is shown in Table 4.

**4.1.4 MLP**

The results from the MLP model show that on training an accuracy of 98% can be achieved and on testing 89% accuracy can be achieved with similar results on mean accuracy after 10-fold stratified cross-validation, as shown in Table 5. The MLP model training and loss history can be visualized through Figure 16.

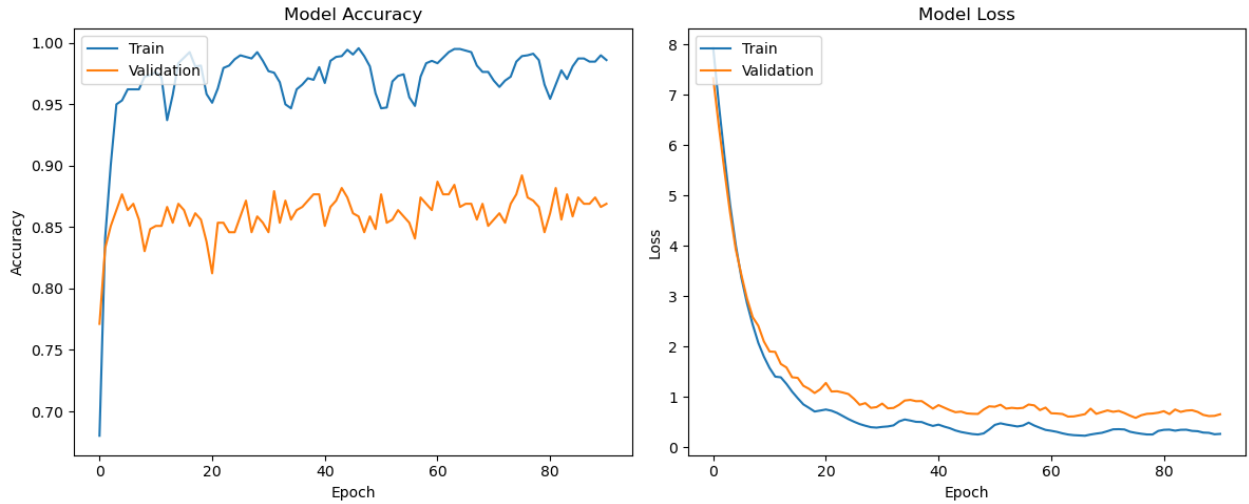

**Fig 16.** Model history plots for the MLP model

The approximately 98% training accuracy and the constantly low training loss show that the model has performed well, fitting the training data almost perfectly. The training accuracy improves rapidly during the initial epochs and stabilizes near 1.0, while the validation accuracy levels off around 0.85–0.88 with minor fluctuations. Similarly, the training loss decreases consistently and stabilizes below 0.5, while the validation loss reduces initially and then stabilizes around 1.0. This behavior indicates that the model effectively learns from the training data, while the validation results highlight a difference in performance that could be addressed with further optimization techniques to enhance generalization.

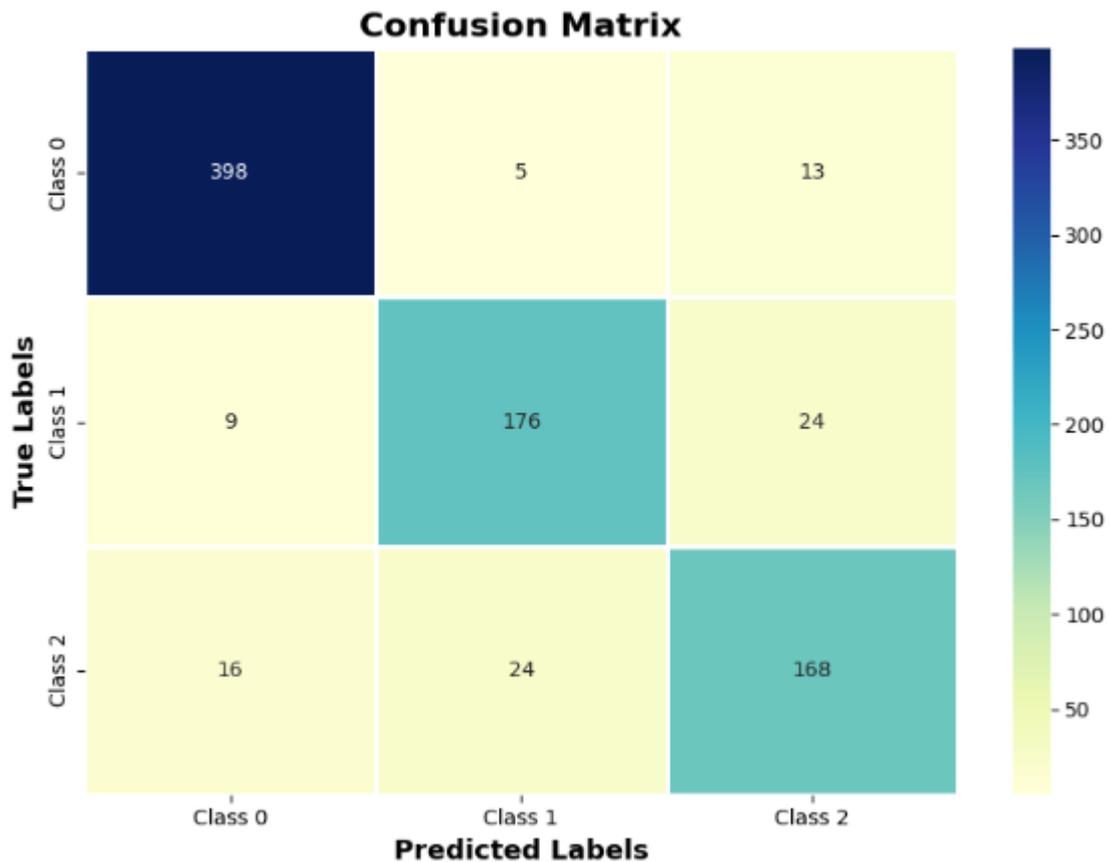

**Fig 17.** Confusion matrix plot for MLP algorithm

The confusion matrix plot for hybrid MLP is shown in Figure 17. The confusion matrix plot as shown in figure shows the very good results with 398 instances out of 416 total predicted correctly for class '0' that is rest state, similarly, out of total of 209 instances 176 were predicted correctly as class '1' is keypress 'd' action and finally, out of total 208 instances 168 were predicted correctly as class '2' which is keypress 'l' action. The hyperparameters for the model are tabulated in Table 4.

**4.1.5 Bi-Directional GRU-Attention model results**

The BiGRU-Attention model was trained and evaluated to achieve strong classification performance on the EEG dataset. The model attained an impressive training accuracy of 98% and a validation accuracy of 90% as shown in Figure 18. While the training curve shows consistent improvement, the validation curve demonstrates stable performance without overfitting, as evidenced by the convergence of both accuracy and loss metrics. The model's loss values also decreased steadily during training, reaching near-optimal levels, as depicted in the loss plot in figure 18.

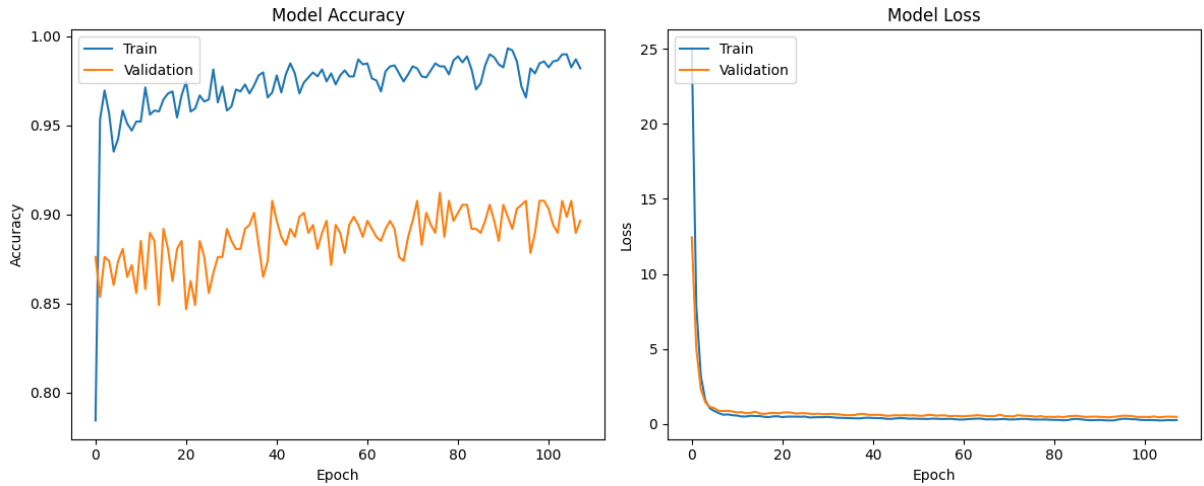

**Fig. 18.** Accuracy and Loss history plot for BiGRU-Attention hybrid model

The classification report reveals strong performance across all classes, with an overall accuracy of 90% as shown in Figure 17. For Class '0', the model correctly classified 263 out of 278 instances, with only 15 misclassifications (5 as Class '1' and 10 as Class '2'), demonstrating high reliability. Class '1' achieved 119 correct classifications, with minor misclassifications of 8 instances as Class '0' and 12 instances as Class '2'. Similarly, Class '2' showed solid performance with 116 correct classifications, while 10 and 13 instances were misclassified as Class '0' and Class '1', respectively. The macro average and weighted average precision, recall, and F1-scores are all 0.90, indicating balanced and consistent performance across all classes. The BiGRU-Attention model's ability to dynamically focus on critical temporal features through its attention mechanism contributed to its robust classification performance, making it well-suited for EEG-based sequential data analysis tasks. These results emphasize the model's stability and effectiveness in multi-class classification scenarios. The best hyperparameters for the model development are tabulated in Table 4 and the confusion matrix plot is shown in Figure 19.

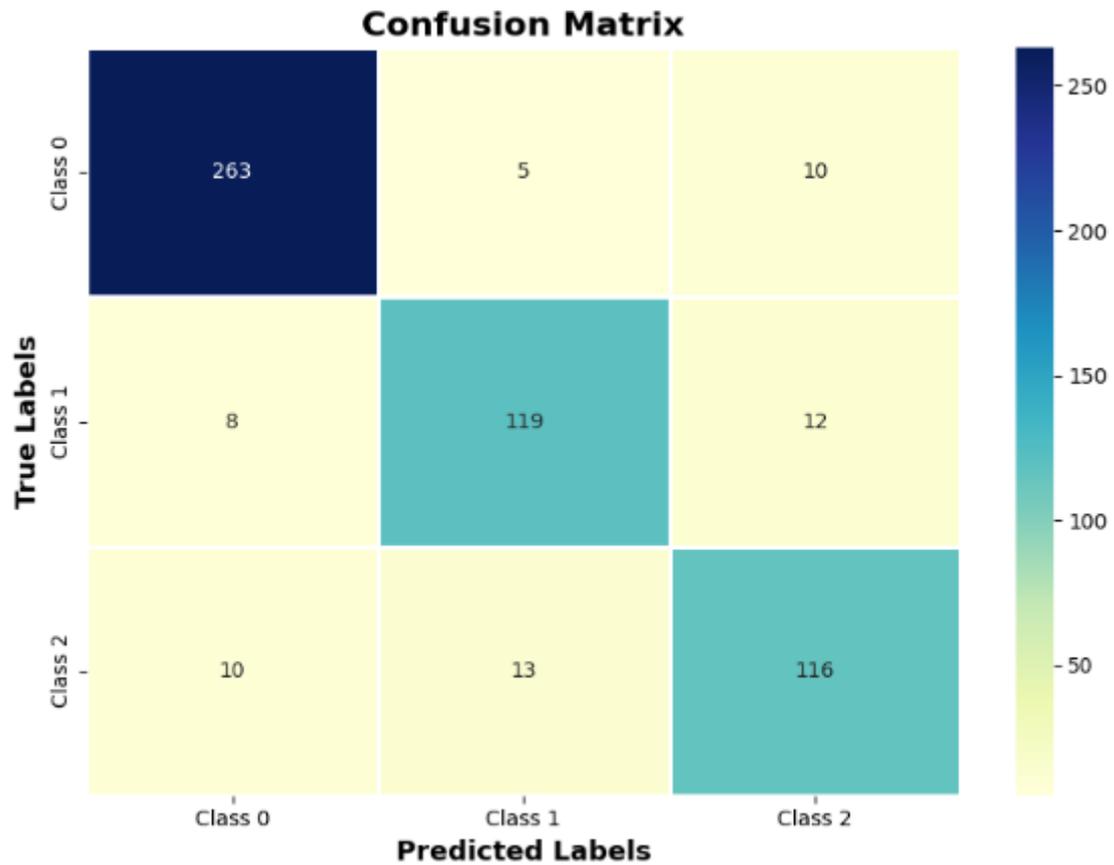

**Fig 19.** Confusion matrix plot for hybrid BiGRU Attention model

**Baseline model**

**4.1.6 Transformer Model**

The baseline transformer-based model architecture was evaluated with early stopping implemented using patience of 15 epochs to prevent overfitting. The accuracy plot shows a consistent improvement in training accuracy, stabilizing at around 95% after approximately 30 epochs, demonstrating the model's capacity to learn effectively from the training data. The validation accuracy converges to a range of 85% to 88%, indicating good generalization to unseen data with a reasonable gap from the training accuracy. The loss plot highlights a rapid decline in both training and validation loss during the initial epochs, stabilizing after approximately 15 epochs. The close alignment of training and validation loss curves, combined with the use of early stopping, ensures that the model avoids overfitting and maintains strong generalization. These results validate the robustness of the model in performing the classification task while effectively balancing training and validation performance. The model history plot and confusion matrix is given in Figure 20 respectively.

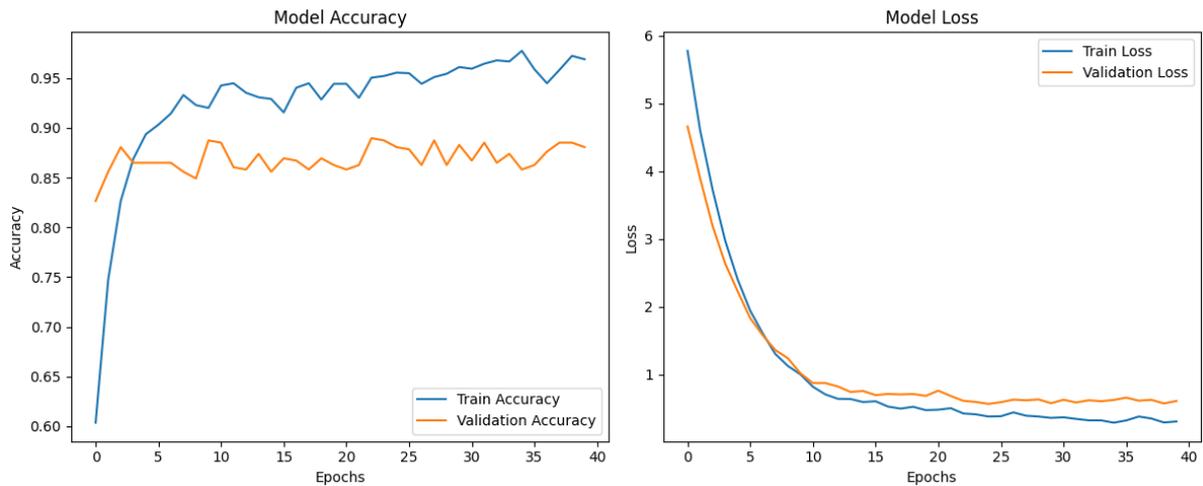

**Fig 20.** Model history plots for Transformer-based algorithm

The confusion matrix plot as seen in Figure 21 depicts that model performance is consistent and performs well on the dataset for class '0' while for classes '1' and '2' it's not the best result obtained. The model predicts the testing data as class '0' 252 times correctly 118 times correct prediction are made for class '1' data and 111 times correct prediction are made for class '2'.

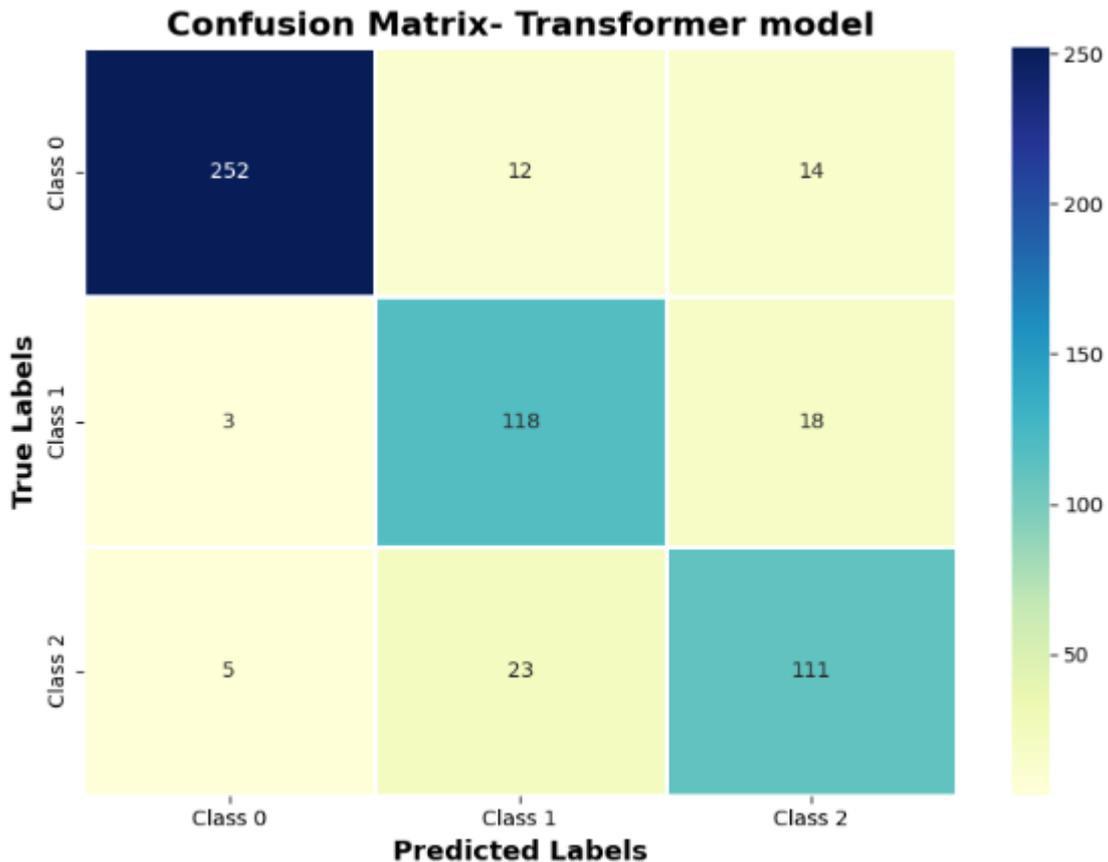

**Fig 21.** Confusion matrix plot for Transformer-based algorithm

### 4.1.7 CTNeT Model

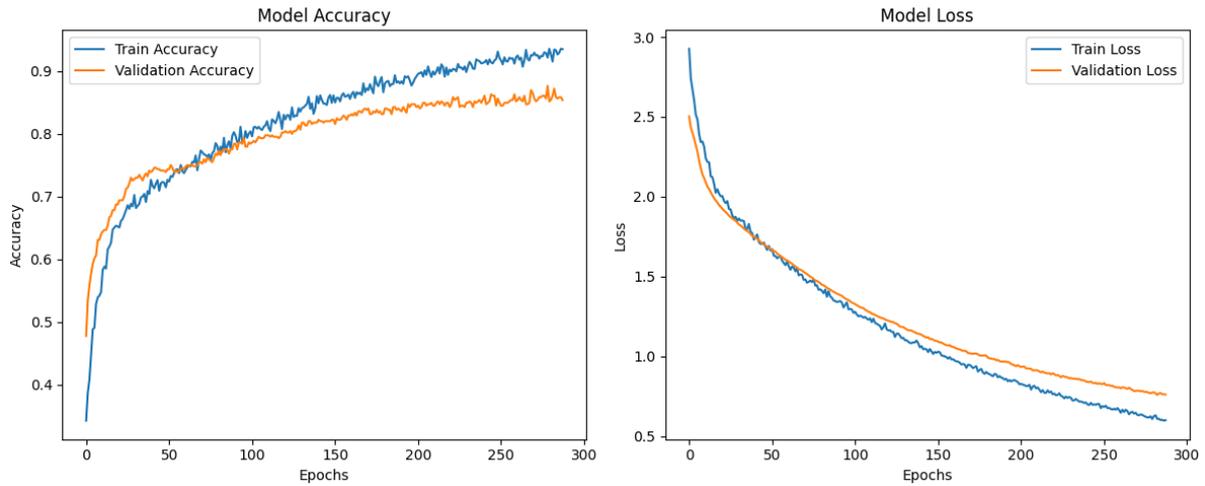

**Fig 22.** Model history plots for the CTNeT algorithm

The CTNeT model was evaluated early, showing a steady improvement in training accuracy, which reached approximately 92%, while validation accuracy stabilized between 83% and 85%. The loss plot indicates a smooth decline in both training and validation loss, with training loss nearing 0.5 and validation loss stabilizing around 0.8. The close alignment of training and validation curves highlights effective learning and minimal overfitting, demonstrating the model's robustness and stability in handling the classification task. The model history plot is depicted in Figure 22.

The confusion matrix plot for the CTNet model is shown in Figure 23. The results indicate strong performance, with 269 out of 278 instances correctly classified as class '0'. Similarly, for class '1', the model correctly predicted 113 instances out of 139, while for class '2', 107 out of 138 instances were accurately classified. This demonstrates the model's ability to distinguish between the three classes effectively.

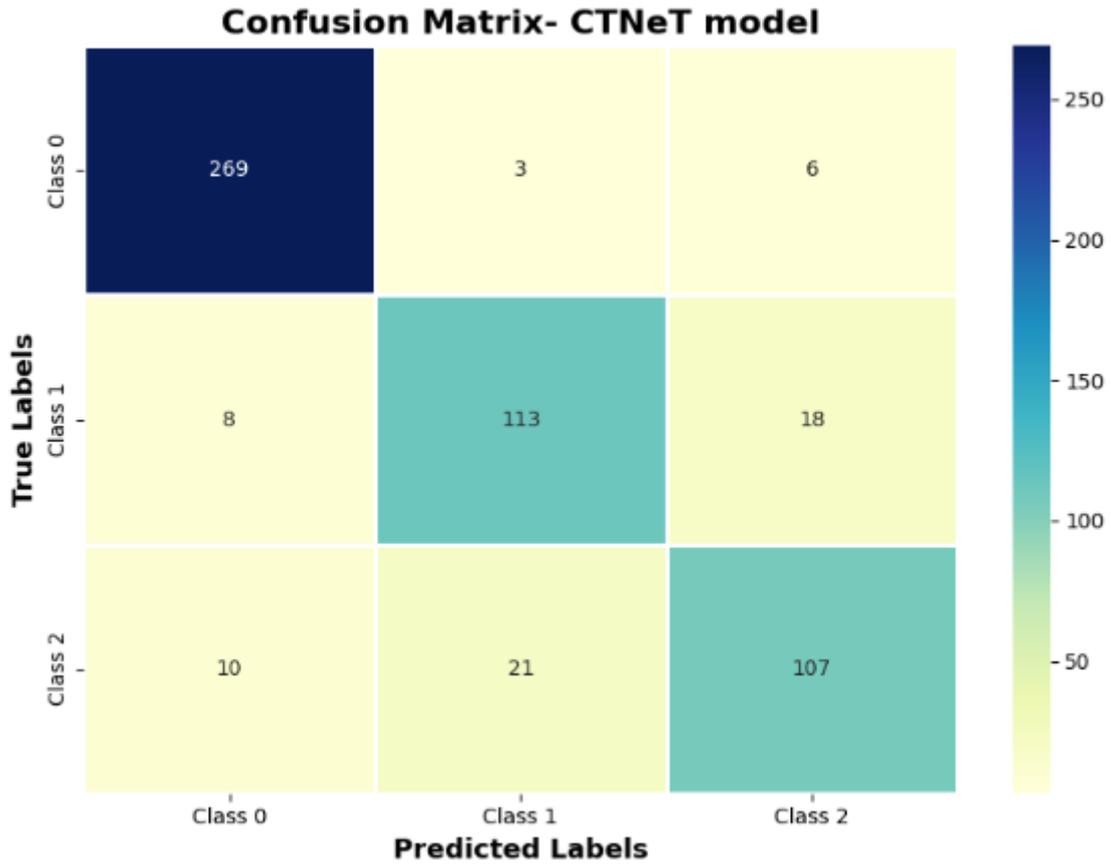

**Fig 23.** Confusion matrix plot for CTNeT algorithm

### 4.1.8 EEGNet

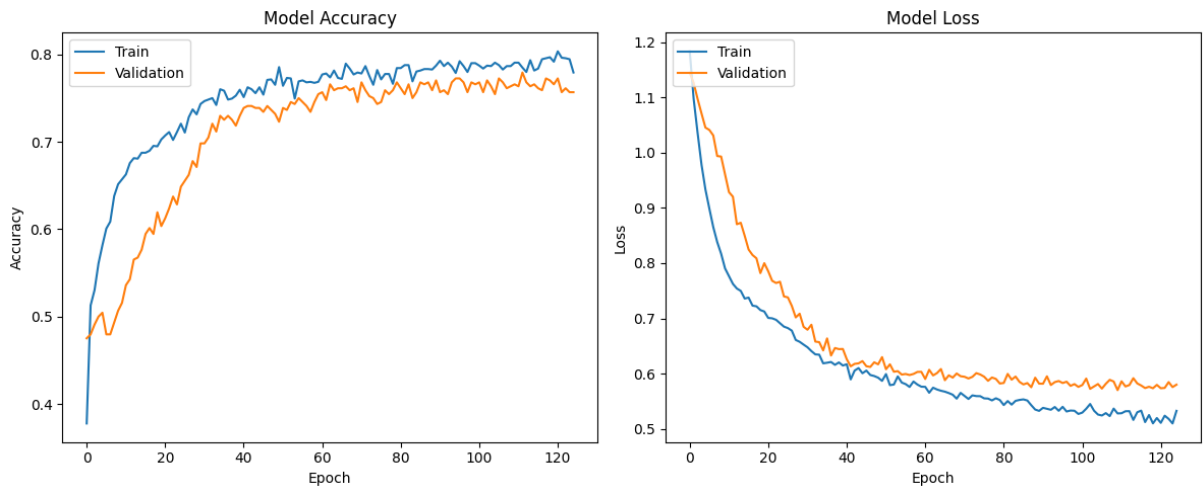

**Fig 24.** Model history plots for EEGNet algorithm

The EEGNet model demonstrated a steady improvement in training accuracy, reaching approximately 80%, while validation accuracy stabilized between 73% and 76%. The loss plot shows a smooth decline in both training and validation loss, with training loss nearing

0.5 and validation loss stabilizing around 0.6–0.7 as shown in Figure 24. The close alignment between training and validation curves highlights effective learning with minimal overfitting, demonstrating the model's robustness and stability in handling the classification task. These results indicate that EEGNet balances model complexity and generalization effectively.

The confusion matrix plot for the EEGNet model is shown in Figure 25. The results indicate strong classification performance across the three classes. For class '0', the model correctly classified 253 out of 278 instances, showcasing high precision for this class. Similarly, for class '1', 81 out of 139 instances were accurately classified, while for class '2', the model correctly predicted 90 out of 139 instances. These results demonstrate the EEGNet model's effectiveness in distinguishing between the three classes, though some misclassifications are observed, particularly between classes '1' and '2'.

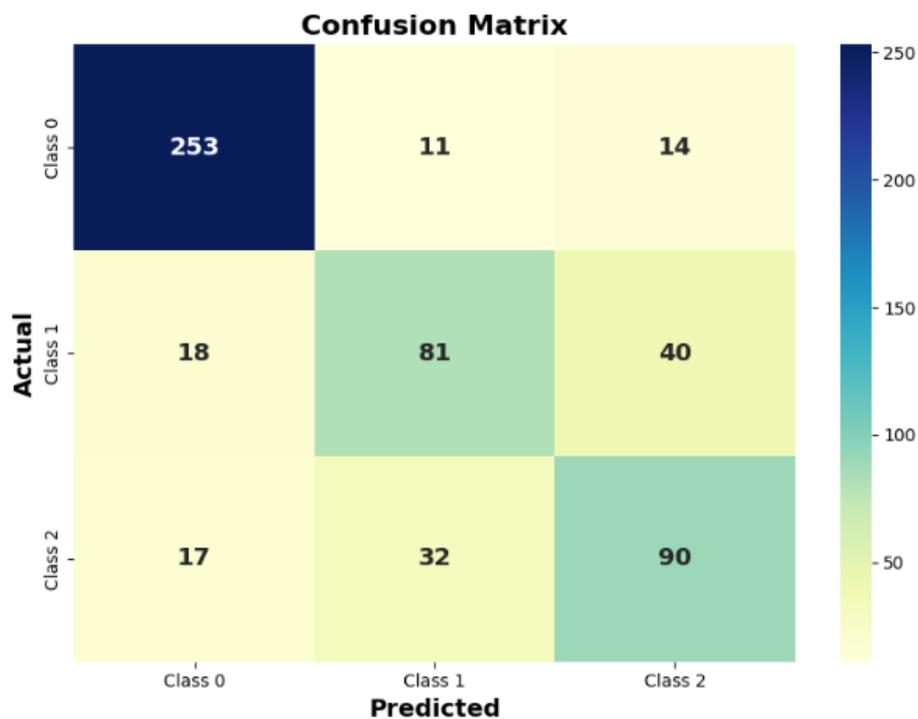

**Fig 25** Confusion matrix plot for EEGNet model

### 4.2. GUI output

The anticipated outcome of this BCI project is a reliable and accurate virtual keyboard system that utilizes EEG signals to predict and simulate voluntary key presses. The system employs a deep learning-based ML model trained to classify EEG data into three categories: "resting stage" (0), "d" key press (1), and "l" key press (2). The numpy array of testing data is uploaded in the '.npy' extension format to test the predictions and simulate brain activity. By integrating the trained model into a Tkinter-based GUI, the system can simulate key presses in real-time based on the user's brain activity. This assistive technology aims to enhance interaction with digital platforms for individuals with neurodegenerative disabilities, providing them with a more effective and accessible means of communication through a brain-controlled virtual keyboard.

On successful testing with our approach, we predict the EEG signal-based classification and perform keypress-based simulation on the GUI, as shown in the Figure 26.

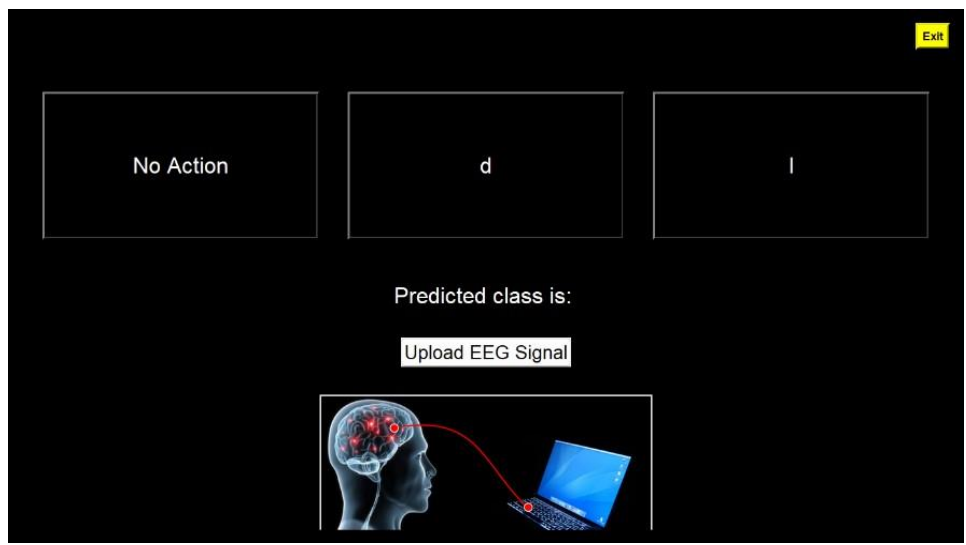

**Fig 26.** Virtual Keyboard GUI

**Table 5:** Stratified cross-validation results of all models

| Fold no. | SVM | NB | CATB | MLP | Transformer-based | CTNeT | EEGNet | BiGRU-Attention Model |
|---|---|---|---|---|---|---|---|---|
| Fold 1 | 0.8993 | 0.7878 | 0.8705 | 0.87 | 0.8777 | 0.8633 | 0.74 | 0.88 |
| Fold 2 | 0.8957 | 0.7950 | 0.8741 | 0.88 | 0.8921 | 0.8417 | 0.77 | 0.90 |
| Fold 3 | 0.9173 | 0.8237 | 0.8885 | 0.88 | 0.8957 | 0.8633 | 0.80 | 0.92 |
| Fold 4 | 0.8705 | 0.7878 | 0.8597 | 0.86 | 0.8633 | 0.8561 | 0.76 | 0.87 |
| Fold 5 | 0.9173 | 0.7986 | 0.8777 | 0.91 | 0.8885 | 0.8741 | 0.81 | 0.95 |
| Fold 6 | 0.8957 | 0.7590 | 0.8345 | 0.88 | 0.8777 | 0.8411 | 0.70 | 0.88 |
| Fold 7 | 0.9314 | 0.8159 | 0.9061 | 0.92 | 0.9206 | 0.8519 | 0.75 | 0.94 |
| Fold 8 | 0.9061 | 0.7906 | 0.8773 | 0.89 | 0.8953 | 0.8628 | 0.74 | 0.90 |
| Fold 9 | 0.8917 | 0.7690 | 0.8412 | 0.87 | .8917 | 0.8916 | 0.74 | 0.89 |
| Fold 10 | 0.9170 | 0.7942 | 0.9097 | 0.89 | 0.8917 | 0.8772 | 0.72 | 0.93 |
| Mean Testing Accuracy | 0.90 | 0.7921 | 0.8739 | 0.89 | 0.8894 | 0.8623 | 0.75 | **0.91** |
| Standard Deviation | 0.016 | 0.018 | 0.023 | 0.02 | 0.014 | 0.015 | 0.03 | 0.02 |

## 4.3. Discussion

We compared our proposed approach with the SVM, GNB, CatBoost, MLP, Transformer-based, CTNeT, and EEGNet. The test accuracy comparison is shown in Table 6, and the classification report comparison is shown in Table 7. For every model, the metrics consist of precision, recall, F1-score, and support for three classes (0, 1, 2). The overall accuracy comparison shows that our proposed model BiGRU-Attention model has the best performance. The SVM and MLP then show the next best model performance.

**Table 6.** Model test accuracy comparison

| Model | Accuracy |
|---|---|
| SVM | 89% |
| GNB | 77% |
| CatBoost | 87% |
| MLP | 89% |
| CTNeT model | 86% |
| Transformer-based | 88% |
| EEGNet | 76% |
| BiGRU-Attention | **90%** |

Evaluating the performance of various models is essential for assuring reliable and consistent results in constructing a virtual keyboard that uses EEG data to categorize key press occurrences. With the highest test accuracy of 90% and highest mean cross-validation accuracy of 91% of the models examined, the Bi-Directional-Attention model outperforms others. With an SD of 0.2 and such high accuracy, proposed model performs well and demonstrates a respectable degree of consistency between runs. The BiGRU-Attention model stands out due to its ability to capture both temporal dependencies and critical features through the integration of BiGRUs and an attention mechanism.

Additionally, MLP and SVM model performs well, achieving a mean cross-validation accuracy of 90% and 89% respectively. SVM and MLP demonstrated high accuracy and consistency, but SVM's reliance on a fixed kernel limits its flexibility in capturing non-linear EEG patterns, while both lacked the dynamic temporal prioritization offered by the BiGRU-Attention mechanism. Transformer-based models and CTNeT, despite their advanced capabilities in handling long-range dependencies and spatial features, exhibited slightly lower accuracies, potentially due to their sensitivity to noise and higher data requirements, underscoring the necessity of architectures like BiGRU-Attention that dynamically prioritize relevant features while efficiently managing noise.

EEGNet, designed for EEG signal processing, achieved 76% accuracy, lower than BiGRU-Attention. Its convolution-based architecture excels at spatial, and spectral filtering but lacks explicit temporal modeling, which is crucial for capturing dynamic patterns in voluntary hand movement classification. This limitation, combined with its sensitivity to noise in multi-class datasets, underscores the advantage of models like BiGRU-Attention, which integrate temporal dependencies and attention mechanisms for improved performance.

The BiGRU-Attention, SVM, MLP models generally show high accuracy and reliability, which makes them excellent choices for categorizing important press events in the creation of EEG-based virtual keyboards. The application's exact need for consistency and dependability determines which of these models to choose.

**Table 7.** Overall model classification report comparison

| Model | Class | Precision | Recall | F1-Score | Support |
|---|---|---|---|---|---|
| SVM | 0 | 0.95 | 0.94 | 0.94 | 278 |
|  | 1 | 0.87 | 0.84 | 0.85 | 139 |
|  | 2 | 0.82 | 0.85 | 0.83 | 139 |
| SVM Accuracy |  |  |  | 0.89 | 556 |
| SVM Macro Avg |  | 0.88 | 0.88 | 0.88 | 556 |
| SVM Weighted Avg |  | 0.89 | 0.89 | 0.89 | 556 |
| GNB | 0 | 0.86 | 0.80 | 0.87 | 278 |
|  | 1 | 0.70 | 0.71 | 0.71 | 139 |
|  | 2 | 0.64 | 0.57 | 0.60 | 139 |
| GNB Accuracy |  |  |  | 0.77 | 556 |
| GNB Macro Avg |  | 0.74 | 0.74 | 0.74 | 556 |
| GNB Weighted Avg |  | 0.77 | 077 | 0.77 | 556 |
| CatBoost model | 0 | 0.92 | 0.97 | 0.94 | 278 |
|  | 1 | 0.83 | 0.81 | 0.82 | 139 |
|  | 2 | 0.86 | 0.78 | 0.82 | 139 |
| CatBoost Accuracy |  |  |  | 0.88 | 556 |
| CatBoost Macro Avg |  | 0.87 | 0.85 | 0.86 | 556 |
| CatBoost Weighted Avg |  | 0.88 | 0.88 | 0.88 | 556 |
| MLP | 0 | 0.94 | 0.96 | 0.95 | 416 |
|  | 1 | 0.86 | 0.84 | 0.85 | 209 |
|  | 2 | 0.82 | 0.81 | 0.81 | 208 |
| MLP Accuracy |  |  |  | 0.89 | 833 |

| Model | Class | Precision | Recall | F1 | Support |
|---|---|---|---|---|---|
| MLP Macro Avg | | 0.87 | 0.87 | 0.87 | 833 |
| MLP Weighted Avg | | 0.89 | 0.89 | 0.89 | 833 |
| Transformer-based | 0 | 0.97 | 0.91 | 0.94 | 278 |
| | 1 | 0.77 | 0.85 | 0.81 | 139 |
| | 2 | 0.78 | 0.80 | 0.79 | 139 |
| Transformer-based Accuracy | | | | 0.87 | 556 |
| Transformer-based Macro Avg | | 0.84 | 0.85 | 0.84 | 556 |
| Transformer-based Weighted Avg | | 0.87 | 0.87 | 0.87 | 556 |
| CTNeT | 0 | 0.94 | 0.97 | 0.95 | 278 |
| | 1 | 0.79 | 0.81 | 0.79 | 139 |
| | 2 | 0.80 | 0.74 | 0.77 | 139 |
| CTNeT Accuracy | | | | 0.87 | 556 |
| CTNeT Macro Avg | | 0.84 | 0.84 | 0.84 | 556 |
| CTNeT Weighted Avg | | 0.87 | 0.87 | 0.87 | 556 |
| EEGNet | 0 | 0.88 | 0.91 | 0.89 | 278 |
| | 1 | 0.65 | 0.58 | 0.62 | 139 |
| | 2 | 0.62 | 0.65 | 0.64 | 139 |
| EEGNet Accuracy | | | | 0.76 | 556 |
| EEGNet Macro Avg | | 0.72 | 0.71 | 0.72 | 556 |
| EEGNet Weighted Avg | | 0.76 | 0.76 | 0.76 | 556 |
| BiGRU-Attention | 0 | 0.94 | 0.96 | 0.95 | 278 |
| | 1 | 0.89 | 0.85 | 0.87 | 139 |
| | 2 | 0.84 | 0.83 | 0.84 | 139 |
| BiGRU-Attention Accuracy | | | | 0.90 | 556 |
| BiGRU-Attention Macro Avg | | 0.89 | 0.88 | 0.88 | 556 |
| BiGRU-Attention Weighted Avg | | 0.90 | 0.90 | 0.90 | 556 |

This research study's accomplishment is comparable to the state-of-the-art research outcomes of an EEG-based keyboard [27], where an SVM model was created for an RGB keyboard for distinct uses. Our method outperformed the previous outcome from SVM, which in this study obtained an 89% test accuracy and a 90% mean accuracy in 10-fold stratified cross-validation.

In recent years, significant progress has been made in the field of BMIs, particularly with advancements that connect cerebral activity to mechanical movement. Early studies demonstrated the feasibility of using electroencephalogram (EEG) signals to control wheelchairs [17], showcasing the potential of non-invasive brainwave monitoring to enhance mobility for individuals with physical disabilities. These developments paved the way for more sophisticated applications, such as EEG-controlled robotic arms for lift-and-grasp tasks [19] and hybrid EEG-EOG-based virtual keyboards too, which allowed users to interact with computers using combined brainwave and eye movement data. Similarly, hybrid BCI systems have been gaining a great enhancement in robotic technology and assistive technology development with EEG [3, 25].

Despite these advancements, EEG-based systems still face challenges, particularly in virtual keyboard applications, where accuracy and user experience remain limited due to signal noise and low resolution [1], [3], [15]. Technologies such as NeuroSky and Emotiv EPOC headsets [26], along with those that integrate blink activity [22, 23], have shown potential in systems for assistive applications. Developments of technologies such as AI and embedded systems with computer vision-based approaches [22], [24] have shown promise for ALS patient assistance and robotic control. Similarly, advancements in mouse cursor control have progressed with EEG after 1991 [13] and enhanced the field of BCI assistive technologies. Integrating EEG with AI technologies, including computer vision [14], offers new possibilities for improving system performance. However, further research is required to address sensor limitations, optimize EEG signal processing, and enhance the robustness of these systems.

Table 8 provides a comparative overview of prior methodologies and their results, highlighting key achievements and identifying critical areas for future work in BCI-based systems.

**Table 8:** Comparison with previous works in related areas
:

| Study | Objective | Key Findings | Reference |
|---|---|---|---|
| RSVP Keyboard System | To aid letter selection during brain-typing using RSVP and language models | Accurate letter selection in a single or few trials | [2] |

| EEG-EOG Hybrid BCI System | Integration of EEG and EOG for virtual keyboard control | EOG traces extracted from EEG signals and treated as an additional input | [3] |
|---|---|---|---|
| EEG-based Game Control | Combining traditional controls with brain signals in games | Identification of attention levels using EEG signals to control 3D environments | [4] |
| Four States BCI for Neurodegenerative Patients | EEG-based BCI for controlling intelligent devices | Achieved 92.5% average classification accuracy | [5] |
| Development of a brain-controlled mobile robot using alpha brainwaves to assist individuals with neuromuscular disorders | Synchronous, EEG-based BCI system detecting eye blinks, with filtered signals processed by a neural network for robotic guidance | The system achieved 92.1% accuracy in controlling robotic movements (forward, backward, left, right) during experiments with 12 subjects. | [7] |
| BCI System Using EEG and MPG Signals | EEG and MPG signals used to control a wheelchair via RF | Control based on concentration and eye blinking intensity | [8] |
| EEG and Microstate Analysis in ALS Patients | EEG spectral band power analysis in ALS patients | Correlation between EEG band power and disease progression | [11] |
| SSVEP-EOG Hybrid Speller | Integrates SSVEP and EOG for speller system | Average classification accuracy of 94.16% | [12] |
| Communication and control system for individuals with severe motor deficits using mu rhythms. | Training subjects to modulate 8–12 Hz mu rhythm amplitudes to control cursor movement via frequency analysis and | Subjects achieved accurate 1-dimensional cursor control with personalized parameters, with target acquisition in ~3 seconds. | [13] |

| | distribution-based parameterization | | |
|---|---|---|---|
| BCI for ALS Communication via Smartphone | Enable ALS patients to communicate through a smartphone using BCI | Provides a communication interface for ALS patients | [14] |
| Control Tool using Neurosky Mindwave | Uses frontal lobe signals for PC control via BCI | Achieved typing proficiency with 1.55-1.8 WPM | [15] |
| EEG-based Wheelchair Control System | Brain-controlled wheelchair using NeuroSky Mindwave | Portable BCI system for wheelchair movement | [18] |
| Systematic Review of MI-BCI for Wheelchair Control | Overview of MI-BCI applications for wheelchair mobility | Highlights algorithm analysis and classification methods | [19] |
| Triple RSVP Speller | High ITR and accuracy speller using RSVP | Achieved 0.790 accuracy and 20.259 bit/min ITR | [20] |
| Brain-Controlled Robotic Arm | Control of a robotic arm using EEG-based BCI | Achieved high accuracy in multi-DoF tasks | [21] |
| Blink Recognition and User-triggered Actions | Real-time communication via blink detection and actions | Demonstrated accurate blink recognition and timely user actions | [22] |
| EEG-based Brain-Controlled Wheelchair | Wheelchair controlled by brainwaves using NeuroSky | Movement based on concentration levels and blinks | [23] |
| Robotic Vision System via GUI | Robotic arm control using a GUI application | Users specify actions via an interactive GUI | [24] |
| Robotic Arm Control with BCI | Brain-controlled robotic arm with EEG | Successful control of a robotic arm by modulating brain rhythms | [25] |

| BCI-based Virtual Keyboard with eMotiv EPOC | BCI virtual keyboard for motor-disabled users | Suggested improvements for virtual keyboard effectiveness | [26] |
|---|---|---|---|
| Virtual Keyboard using EEG Signals | QWERTY virtual keyboard with EEG control | Achieved 89.7% accuracy with 6.4 CPM | [27] |
| Our work | Utilization of hybrid ML models for making a virtual keyboard based on right and left-hand voluntary movement, combined with the implementation of a Tkinter-based virtual keyboard interface | Achieved 90% best test accuracy on BiGRU-Attention as compared with other ML, Transformer, and Hybrid models. Finally, BCI-based voluntary hand movements-based testing data-based simulating of a virtual TKinter-designed keyboard is possible for predicting 3 actions that include 2 keypresses and a resting state of a person. | |

## 5. CONCLUSION

In this work, we developed an EEG-based virtual keyboard utilizing a BiGRU-Attention hybrid deep learning approach to classify voluntary right- and left-hand movements. The proposed model achieved a test accuracy of 90% with 91% mean cross-validation accuracy, surpassing traditional ML methods like SVM, GNB, and CatBoost, as well as advanced architectures such as transformer-based, hybrid and EEGNet models, establishing its reliability and robustness. The integration of this model into a real-time GUI further validates its practicality, providing an efficient and user-friendly assistive technology solution for individuals with neurodegenerative conditions.

By addressing significant gaps in EEG signal classification and BCI system development, this work demonstrates the transformative potential of deep learning in assistive technologies, particularly in the development of the virtual keyboard. The findings establish a strong basis for future research, including the exploration of multi-class classification, the integration of additional EEG signal modalities, and broader applications in human-computer interaction. This work not only empowers individuals with motor impairments but also advances the frontier of BCI innovation, paving the way for scalable and robust solutions in accessibility technology.